\documentclass[12pt,leqno, oneside]{amsart}

\usepackage{amssymb,amsmath,amsthm, nccmath}
\usepackage[latin1,utf8]{inputenc}
\usepackage{graphicx,color, textcomp}
\usepackage[dvipsnames]{xcolor}
\usepackage{stackrel}
\usepackage{mathtools}
\usepackage[toc,page]{appendix}
\usepackage{comment}

\usepackage[left=2cm,top=2.5cm,bottom=2.5cm,right=2cm]{geometry}
\usepackage[english]{babel}
\everymath{\displaystyle}
\counterwithin{figure}{section}
\mathtoolsset{showonlyrefs}
\usepackage{tikz-cd} 
\usepackage[percent]{overpic}

\def \R {{\mathbb R}}
\def \N {{\mathbb N}}
\def \Z {{\mathbb Z}}

\def \Q {{\mathbb Q}}

\newcommand {\beqn} {\begin{equation*}}
	\newcommand {\eeqn}	{\end{equation*}}
\newcommand {\beq} {\begin{equation}}
	\newcommand {\eeq}	{\end{equation}}

\newtheorem{theorem}{Theorem}[section]

\newtheorem{definition}[theorem]{Definition}
\newtheorem{proposition}[theorem]{Proposition}

\newtheorem{remark}[theorem]{Remark}

\numberwithin{equation}{section}

\title[Analytical methods in Celestial Mechanics]{Analytical methods in Celestial Mechanics: statellites' stability and galactic billiards}
\author{Irene De Blasi}

\address{University of Turin, Department of Mathematics, via Carlo Alberto 10, Turin, Italy}

\date{\today} 
\thanks{Supported by Italian Research Center on High Performance Computing Big
	Data and Quantum Computing (ICSC), project funded by European Union -
	NextGenerationEU - and National Recovery and Resilience Plan (NRRP) -
	Mission 4 Component 2. Spoke 3, Astrophysics and Cosmos Observations.\\
	Work partially supported by INdAM group G.N.A.M.P.A. \\
	The author has no relevant financial or non-financial interests to disclose.}


\makeatletter
\@namedef{subjclassname@2020}{\textup{2020} Mathematics Subject Classification}
\makeatother


\begin{document}
	
\maketitle
\begin{abstract}
	In this paper, two models of interest for Celestial Mechanics are presented and analysed, using both analytic and numerical techniques, from the point of view of the possible presence of regular and/or chaotic motion, as well as the stability of the considered orbits. \\
	The first model, presented in a Hamiltonian formalism, can be used to describe the motion of a satellite around the Earth, taking into account both the non-spherical shape of our planet and the third-body gravitational influence of Sun and Moon. Using semi-analytical techniques coming from Normal Form and Nekhoroshev theories it is possible to provide stability estimates for the orbital elements of its geocentric motion. \\
	The second dynamical system presented can be used as a simplified model to describe the motion of a particle in an elliptic galaxy having a central massive core, and is constructed as a \textit{refraction billiard} where an inner dynamics, induced by a Keplerian potential, is coupled with an external one, where a harmonic oscillator-type potential is considered. The investigation of the dynamics is carried on by using tools of ODEs' theory and is focused on studying the trajectories' properties in terms of periodicity, stability and, possibly, chaoticity. 
\end{abstract}
\section{Introduction}\label{sec:intro}
Phenomena involving the motion of Celestial bodies, either on a planetary or a galactic scale, are often characterised by a complex behaviour, whose accurate study requires the use of different tools, like numerical integration, analytical study, direct observations and much more (\cite{BarCanTeranisotropic, celletti2017dynamical,BolNeg,Kn2002,baldoma2022breakdown}). Analytical techniques represent, whenever applicable, useful strategies to study some of the main properties of the orbits in a gravitational system, especially in terms of \textit{long-term dynamics}, providing results which are \textit{rigorous}, as they follow form precise mathematical statements, and often \textit{general}, in the sense that they potentially hold for a large set of trajectories (equivalently, of initial conditions). \\
Along with purely analytical techniques, in some circumstances a mixed approach which includes numerics as well is possible: this is what happens, for example, when theoretical results are compared with simulations and observations, or in the case of \textit{semi-analytical} approaches. Generally speaking, such expression refers to a class of methods where rigorous mathematical theorems are applied to numerically computed quantities (for example, within a Hamiltonian framework, to functions expressed through a truncated Taylor expansion, cfr. Section \ref{sec:satelliti}) \\
This paper aims to illustrate the potential of such techniques, either analytic or semi-analytic, presenting the dynamical investigation of two models describing the geocentric motion of an object around the Earth and the trajectories of a body inside an elliptic galaxy with a massive core.
In both cases, the motion of our test particle is influenced by the gravitational attraction of a variety of different mass distributions, depending on the model itself: as expected, the resulting dynamics is quite complex, and our main objective is to study its properties for long (possibly infinite) time scales. \\
The issue of the long-term stability in geocentric motions is the core topic of Section \ref{sec:satelliti}, where a point-mass particle subjected to the attraction of the
(non-spherical) Earth, Sun and Moon is taken into account. In general, the main question we try to answer is \textit{for how long} it is possible to control the variation in the orbital elements (semimajor axis $a$, eccentricity $e$, inclination $i$) of our object, considering different  initial conditions and, in particular, for different altitude regimes (we will use the classical distinction between NEO, MEO and GEO distances). Producing stability estimates for orbiting bodies at different distances from our planet's surface is a key problem in Celestial Mechanics, which finds its application to many different cases of practical interest. In particular, this problem is crucial when dealing with the wide and varied class of the object orbiting around the Earth, from satellites to microscopical space debris: in view of their large overall number and the collision hazard (cfr. \cite{report2022}), the effort in predicting as well as possible their long-time behaviour has involved a remarkable community of mathematicians and astronomers (see for example \cite{celletti2021reconnecting, celletti2022proper, daquin2016dynamical, nie2021long, Shute}, and, for a survey on the possible methods, \cite{cellettiIAU}).  \\
Following the vast literature on the subject, the satellites' dynamics is formalised within a Hamiltonian setting via the so-called \textit{lunisolar Hamiltonian} 
\begin{equation*}
		\mathcal H(\underline r,\underline{\dot r}; t) =  \frac{|\underline{\dot{r}}|^2}{2} + \mathcal H_E(\underline r)+ \mathcal H_S(\underline r; t)+ \mathcal H_M(\underline r; t), 
\end{equation*}
where $\underline r, \underline{\dot{r}}$ are the position and velocity vector of our satellite in a suitable reference system, while the three potential parts refers to Earth's geopotential up to $J_2$-term (see Section \ref{sec:model} and \cite{kaula1966theory} for details) and Sun's and Moon's gravitational attractions treated as third-body perturbations. \\
As already anticipated, the techniques we used to produce stability estimates fall into the category of the semi-analytical methods, since $\mathcal H$, opportunely treated, and expressed in its \textit{secular form} (namely, averaged over the fast motions, see Section \ref{sec:model}) can be written as a truncated Taylor expansion whose coefficients are computed numerically. \\
From a mathematical point of view, we propose two different methods to produce stability estimates, holding in different regimes and based on different analytical results. \\
With the first strategy, we provide stability estimates for the quantity $\mathcal I=\sqrt{\mu_E a}\sqrt{1-e^2}(1-\cos{i})$ in quasi circular and quasi equatorial orbits, computing an upper bound for the time up to which the variation of such quantity remains bounded within a certain range. The technique used to produce such estimates (see also \cite{steichen1997long}) is based on the application of a \textit{normal form} algorithm: in short, and postponing the complete description to Section \ref{ssec:primo_art}, we use canonical transformations in action-angle coordinates to reduce the Hamiltonian describing the satellites' motion to the form 
\begin{equation}\label{eq:ham generale}
	\mathcal H=h_0+h_1,  
\end{equation}
where $h_0$ admits $\mathcal I$ as a first integral and the size of $h_1$ is so small that the overall dynamics can be considered a \textit{perturbation} of the one induced by $h_0$. The stability of $\mathcal I$ along the trajectories can be then deduced from the size of $h_1$. The stability results, holding for small values of the eccentricity and inclination and five different values for the semimajor axis, corresponding to NEO, MEO, GEO region and beyond, are shown in Section \ref{ssec:primo_art}, and precisely in Table \ref{tab:stab gls}. In short, one can say that  the numerically computed stability times are extremely long, of the order of $10^{4}$ years even in the worst case, corresponding to the farthest objects, although a worsening, due to the strengthening of the influence of Sun's and Moon's attraction, is evident beyond GEO region. \\
As for the second method, which is the subject of the analysis carried on in Section \ref{ssec:sec_art}, it is based on Nekhoroshev theorem on exponential stability estimates (see \cite{nekhoroshev1977exponential}), and allows to cover a larger domain in eccentricity and inclination for satellites in MEO, and in particular for distances (in terms of semimajor axis) between $11\ 000\ km$ and $19\ 000\ km$.  Nekhoroshev theorem has already been used in some problems coming from Celestial Mechanics, like for example in the model of the Trojan asteroids (\cite{giorgilliskokos}) and in the three-body problem (\cite{cellettiferrara, cellettigiorgilli}), and applies again to the case of a \textit{quasi-integrable} Hamiltonian. Given an Hamiltonian function as in Eq. \eqref{eq:ham generale}, where all the actions are first integrals for the unperturbed dynamics, under suitable conditions it provides stability times in the complete model which are \textit{exponentially long} in the perturbation's size; the hypotheses required involve suitable nondegeneracy conditions on the unperturbed Hamiltonian $h_0$, as well as a smallness condition on the size of $h_1$. \\
In the present work, a nonresonant version of the theorem, which does not apply close to the secular geolunisolar resonances (see \cite{breiter1999lunisolar}), has been used, although a more complete analysis, which covers a wider regime, is possible provided a rigorous analysis of the geometry of resonances of the geolunisolar problem is carried on. The results in terms of stability times, for different values of $a$, $e$ and $i$, are presented in Section \ref{ssec:sec_art}, Figure \ref{fig:nekh stab time}: they are particularly good for low altitudes, and tend to worsen for increasing values of $a$. This phenomenon, that partially depends on the algorithm used to produce our stability estimates, will find an heuristic explanation in Section \ref{ssec:sec_art}. \\

The second model taken into consideration is a simplified model that can be used to carry on a preliminary analysis on the motion of a particle in an elliptic galaxy having a central mass (a Black Hole or, in general, a massive core). 
This kind of motion, especially under the influence of super-massive bodies such as Black Holes, is particularly complex, and having a rigorous and reliable model to describe it would require to take into consideration the anisotropies in the mass distribution inside the galaxy, as well as relativistic effects. Situations of galaxies presenting Black Holes at their centers are quite common in actual galactic systems (see the review \cite{ferrarese2005supermassive}), and it is quite natural, for anyone working in Celestial Mechanics, to ask how the presence of such a large central mass affects the dynamics, as well as whether it could lead to \textit{chaotic phenomena}.  \\
The study we propose Section \ref{sec:biliardi} has been inspired by the work \cite{Delis20152448}, where the central body acts as a Keplerian center and the elliptic distribution of mass produce a harmonic oscillator-type potential. The superimposition of such two potentials leads to the establishment of two different regimes: whenever our test particle is close to the central body, the Keplerian attracting force of the latter is much more intense than the one of the overall galaxy, while the contrary happens whenever the particle is sufficiently far from the Black Hole. When the galaxy's mass distribution is an uniform ellipsoid, one can model its gravitational attraction via a harmonic oscillator-type potential (see \cite{chandrasekhar1967ellipsoidal}), where the frequencies over the three axes of the oscillations depend on the three semiaxes, while, ignoring possible relativistic effects, the potential of the central mass is a classical Keplerian one. In \cite{Delis20152448}, the investigation of the model is carried on by means of a mixed analytical and numerical approach, and evidences of \textit{chaotic behaviour} based on estimates of the corresponding Lyapunov exponents (see \cite{kalapotharakos2008rate}) are shown. \\
In our work (see also \cite{deblasiterracinirefraction, deblasiterraciniOnSome,barutello2023chaotic}), we propose a rigorous analysis which, although substantiated by numerical evidences, relies on a \textit{purely analytical approach:} the price to pay is the necessity to introduce a simpler model, where the superposition of the two potential does not occur anymore. As in \cite{Delis20152448}, we suppose that the distribution of matter in the ellipsoid is constant (except for the central body), and we build the model in such a way that our test particle is either attracted by the central mass or by the overall elliptic mass distribution. In practice, we divide the space into two regions, in each of which one of the two limit regimes identified by \cite{Delis20152448} occurs. In the inner region, representing the \textit{region of influence} of the central mass, only its (Keplerian) attraction is considered; on the contrary, in a second region, exterior, the particle moves only under the influence of an isotropic harmonic oscillator. On the interface that separates these two regions the potential governing the particle's motion is generally discontinuous: to treat such discontinuity, we suppose that every time the particle hits the interface it undergoes a \textit{refraction}, which deflects its velocity by a quantity that depends on the potentials' values in the transition point, as well as the hitting angle. Such refraction law, which in practice is a generalisation of classical Snell's law for light rays, can be interpreted as a limit case for a smooth passage from one potential to the other, where the intermediate region in which the two potentials are superimposed shrinks more and more (from a practical point of view, the paper \cite{Delis20152448} also provides estimates from the distance from the central mass at which the two potentials produce comparable forces). \\
At this stage, we restrict our analysis on a \textit{planar} system, considering one of the three invariant planes of the system identified by the ellipsoid's axes; this assumption will be removed in the next future (\cite{deblasi3D}), where the three dimensional case will be considered.\\
From a mathematical point of view, we construct our model by relying on the well established theory of \textit{mathematical billiards} (see for example \cite{Tabbook} for an extensive survey on the classical theory), and constructing the so-called \textit{galactic refraction billiard}. 
\begin{figure}
	\centering
	\includegraphics[width=0.7\linewidth]{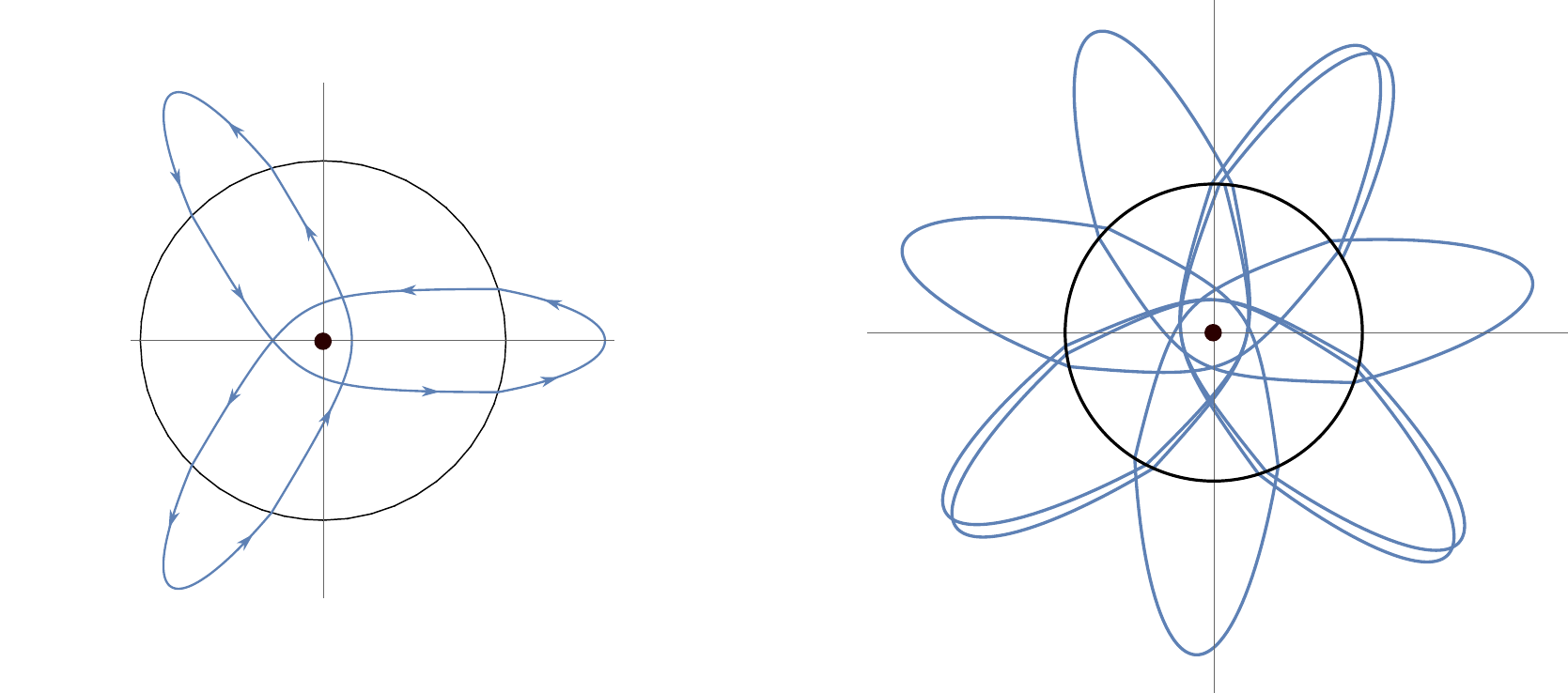}
	\caption{Examples of orbits of refraction galactic billiards. The orbit goes inside and outside the domain, being deflected at every passage through the interface. Left: three-periodic trajectory. Right: quasi-periodic trajectory (figure taken from \cite{deblasiterraciniOnSome}). }
	\label{fig:traj} 
\end{figure}
In the classical case of mathematical billiards, a free particle moves inside a regular domain, following straight lines and bouncing against the boundary with an elastic reflection; deriving the properties of the particle's motion (equilibrium trajectories, existence of periodic orbits, chaotic regimes etc.) is a highly nontrivial problem, which has involved a wide community of mathematicians for at least one century.\\
Within this framework, our billiard can be considered as a variation of the classical case, where the inner mass' domain of influence represents our billiard table, although two important differences have to be highlighted. First of all, the particle can exit from the domain, interacting with its boundary not with a simple reflection, but rather with a refraction that deflects its velocity vector. On the other hand, the presence of the inner and outer gravitational interactions leads to the appearance of two (outer and inner) non constant potential, so that the particle moves through curved geodesics instead of straight lines. Other examples of billiards with potentials, both in the reflective case and in the case of a coupled dynamics, are given in \cite{gasiorek2021dynamics, lerman2021whispering}; a remarkable example is given by Kepler billiards (see \cite{panov1994elliptical, takeuchi2021conformal} and references therein), which, as we will see in Section \ref{ssec:bil terzo_art}, present strong analogies with our model. \\
The results summarised in the present paper regard different aspects of the dynamics of the refraction galactic billiard for different domain's shapes and energy regimes, and have been achieved by using a wide class of tools coming from nonlinear analysis and the general theory of dynamical systems, as well as, sometimes, substantiated by numerical simulations. They will be presented into three main subgroups; first of all, as natural while dealing with a new dynamical system, the problem of existence and stability of equilibrium trajectories is considered. This is the topic of Section \ref{ssec:bil primo_art}, where a particular class of equilibrium orbits, called  \textit{homothetic} and composed by straight lines, is considered. Such trajectories always exist when our domain is convex and smooth, and their linear stability can be studied by relying on the formalism of classical billiards and variational methods. In this framework, nontrivial bifurcation phenomena, occurring for non-circular domains with sufficiently smooth boundary, are shown, both from an analytical and a numerical point of view. \\
After this preliminary (and local) analysis on the trajectory orbits, the investigation becomes more global in Section \ref{ssec:bil secondo_art}: here, the problem of the existence of periodic and quasi-periodic trajectories (see Figure \ref{fig:traj}) is treated. We work in a \textit{quasi-integrable} regime, considering domains whose boundary is close to a circumference, and, as a consequence, can be treated with the powerful tools of perturbation theory (see also \cite{cellettiIAU}, where such concepts are explained in a slightly different framework). In particular, we shall make use of KAM theorem (see \cite{moser1962invariant}), Poincar\'e-Birkhoff and Aubry-Mather theories (see \cite{gole2001symplectic}) to prove that, whenever our domain's boundary is smooth enough and sufficiently close to a circle, then there exists orbits with any rotation number within a certain range (see Theorem \ref{thm:rot} and Eq. \eqref{eq:rot} for the formal definition of rotation number). We stress that our case is not the first application of KAM, Aubry-Mather and Poincar\'e-Birkhoff theories in problems coming from Celestial Mechanics: examples are \cite{celletti2007kam, boscaggin2021periodic}. \\
The landing point of the analysis of the galactic billiards' dynamics, at least in the regime here presented, is included in Section \ref{ssec:bil terzo_art}, where we address the problem of the possible \textit{chaoticity} of the system. In our specific case, evidences of chaotic behaviour are included in both \cite{Delis20152448} and the numerical simulations presented in \cite{deblasiterracinirefraction} and reproduced in Figure \ref{fig:chaos numerica}: these two elements motivated the prosecution of the study in this direction, trying to formally prove the chaoticity of the model. \\
The final result resumes in the detection of a simple geometric condition on the domain's shape, called \textit{admissibility}, that ensures the existence of a topologically chaotic subsystem of the galactic refraction billiard for large enough inner energies (see Theorem \ref{thm:chaos}). Roughly speaking, and postponing the rigorous definition to Section \ref{ssec:bil terzo_art} (and in particular Definition \ref{def:ammissibile}), we say that a domain with smooth boundary is admissible whenever there exist two segments from the Keplerian mass which are orthogonal to the boundary, not antipodal with respect to the origin and such that they are nondegenerate (that is, the hitting point is a strict maximum or minimum of the function distance from the mass restricted to the domain's boundary), see Section \ref{ssec:bil terzo_art}, Figure \ref{fig:ammissibile}.
In practice, admissibility acts as a sufficient condition that, through a particular construction called symbolic dynamics, ensures that, up to restricting to a subset of all the possible initial conditions and choosing a sufficiently large inner energy, our galactic billiard is chaotic.  \\
Also in this case, we stress that our work can be considered as a part of a vast literature, whose aim is to investigate and detect, with different techniques, chaotic systems in Celestial Mechanics, both with a rigorous analytical approach (see for example \cite{Bol2017, baldoma2022breakdown,baldoma2023breakdown}), and with a more numerical point of view \cite{guzzo2023theory,froeschle1997fast}.

\section{Hamiltonian methods for satellites' stability estimates}\label{sec:satelliti}

The current Section summarises the results of \cite{de2021semi, cellettiDeBlasiEft2023nekhoroshev} regarding the long-term stability for bodies orbiting around the Earth, considering, in a Hamiltonian setting, the gravitational attraction of our planet, Sun and Moon. \\
Section \ref{sec:model} describes the Hamiltonian model taken into consideration, including the set of action-angle variables used, with particular attention on their physical meaning in terms of orbital elements. Section \ref{ssec:primo_art} resumes the main ideas behind normal form theory, proposing then the application of such approach to our model to produce stability estimates for eccentricity and inclination, locked in the quasi-integral $\mathcal I=\sqrt{\mu_E a}\sqrt{1-e^2}(1-\cos{i})$, for quasi circular and quasi equatorial orbits. Section \ref{ssec:sec_art} widen the set of the considered initial conditions to more inclined and eccentric orbits within MEO distances: in this case, an approach based on the application of Nekhoroshev theorem is taken into account. Finally, Section \ref{ssec:conclusioni sat} presents some final considerations on the results obtained, comparing the two approaches both in terms of the numerical outcome and theoretical consequences. 

\subsection{The Hamiltonian model}\label{sec:model}

To construct the Hamiltonian function related to the geolunisolar model, let us start by considering a geocentric reference frame in the space, with coordinate axes $x,y \text{ and } z$, where the $(x,y)-$plane corresponds to the Earth's equatorial one and the $x-$axis points towards the line of the equinox. In such framework, the geolunisolar Hamiltonian referred to a point-mass particle of coordinate vector $\underline{r}=(x,y,z)$ can be expressed by\footnote{In the current paper, unless otherwise specified, the norm $|\cdot|$ denotes the Euclidean norm in $\R^N$. }
\begin{equation}\label{eq:hamiltoniana cart}
    \mathcal H(\underline r,\underline{\dot r}; t) =  \frac{|\underline{\dot{r}}|^2}{2} + \mathcal H_E(\underline r)+ \mathcal H_S(\underline r; t)+ \mathcal H_M(\underline r; t), 
\end{equation}
where the potential terms are given as follows: 
\begin{itemize}
    \item the term $\mathcal H_E$ is the Earth's gravitational potential which takes into account the non-spherical shape of our planet; it can be expressed as an expansion in spherical harmonics, as described in \cite{Kaula1962}. In the current model, such expansion is truncated up to the $J_2$-term\footnote{The choice of this truncation is motivated by the fact that this term is in magnitude much bigger with respect to the following ones, see for example \cite{celletti2017dynamical}. A comparison of the current model with the one in which more harmonics are considered is presented in Section \ref{ssec:conclusioni sat}}, giving rise to an expression of the form 
    \begin{equation}\label{eq:ham J2}
        \mathcal H_E(\underline r)=-\frac{\mu_E}{|\underline r|} - J_2 \frac{\mu_E R_E}{|\underline r|^3}\left(\frac{1}{2}-\frac{3 z^2}{2|\underline r|^2}\right), 
    \end{equation}
    where $R_E=6378.14\ km$ and $\mu_E=\mathcal G M_E = 1.52984 \times 10^9\ R_E^3/yr^2$ are respectively the Earth's radius and mass parameter and $J_2=-1082.6261\times 10^{-6}$ is a dimensionless parameter; 
    \item the terms $\mathcal H_S$ and $\mathcal H_M$ refer to the gravitational attraction of Sun and Moon, whose motion in the geocentric reference frame is given respectively by the time-dependent position vectors $\underline r_S(t)=(x_S(t), y_S(t), z_S(t))$ and $\underline r_M(t)=(x_M(t), y_M(t), z_M(t))$. More precisely, one has 
    \begin{equation*}
        \mathcal H_{S\backslash M}(\underline r; t) =  -\mu_{S\backslash M}\left(\frac{1}{|\underline r-\underline r_{S\backslash M}|}-\frac{\underline r\cdot\underline r_{S\backslash M}}{|\underline r_{S\backslash M}|^3}\right), 
    \end{equation*}
where again $\mu_S$ and $\mu_M$ are respectively the mass parameters of Sun and Moon. As for the analytic expression of $\underline r_S$ and $\underline r_M$, one has that both bodies moving around the Earth describing ellipses: the orbital parameters  (inclination $i_0$, semimajor axis $a$ and eccentricity $e$) of Sun are $i_{0S}=23.43^\circ$, $a_S=1.469\times 10^8\ km$ and $e_S=0.0167$, while for the Moon one has  $i_{0M}=i_{0S}$, $a_M=38\ 4748\ km$ and $e_M=0.065$. 
\end{itemize}
From a dynamical point of view, assuming that the Moon lies \textit{on the ecliptic plane} corresponds to neglect the precession of the Lunar node: as will be observed later in Section \ref{ssec:sec_art}, this assumption will have important effects close to the so-called secular lunisolar resonances (see also \cite{breiter1999lunisolar}).  \\
Since the motion of our point-mass particle is a geocentric trajectory, it is  convenient to express the Hamiltonian \ref{eq:hamiltoniana cart} in terms of the particle's orbital elements; such change of variables is performed by expressing, as in \cite{DebCelEft2021satellites} and \cite{murray1999solar}, the coordinates $x,y,z$ (resp. the components $x_{S\backslash M}, y_{S\backslash M}, z_{S\backslash M}$ of $\underline r_{S\backslash M}$) in terms of orbital elements $(a,e,i, M, \omega,\Omega)$ (resp. $a_{S\backslash M},e_{S\backslash M},i_{S\backslash M}, M_{S\backslash M}, \omega_{S\backslash M}$ and $\Omega_{S\backslash M}$), where $a, e$ and $i$ denote respectively the orbit's semimajor axis, eccentricity and inclination, while the angles $M, \omega$ and $\Omega$ are the mean anomaly, the argument of the perigee and the longitude of the nodes. The resulting Hamiltonian, which will still be called $\mathcal H$, is a function of  $(a,e,i,M, \omega, \Omega)$, where where the time dependence is expressed by the mean anomalies $M_S$ and $M_M$ of Sun and Moon.\\
When one in interested to the satellite's \textit{long term} dynamics, the Hamiltonian $\mathcal H$ can be further simplified, removing the dependence on the time, by considering an averaging process over the fast angles of the problem (namely, the three mean anomalies): the result of this averaging is the \textit{secular geolunisolar Hamiltonian}
\begin{equation}\label{eq:ham sec}
\begin{aligned}
    \mathcal H_{sec}(e,i,\omega, \Omega)&=\mathcal H_{E}^{(av)}(e,i,\omega, \Omega)+\mathcal H_{S}^{(av)}(e,i,\omega, \Omega)+\mathcal H_{M}^{(av)}(e,i,\omega, \Omega)\\
    &= \frac{1}{2\pi}\int_0^{2\pi} \mathcal H_E\  dM+\frac{1}{4\pi^2}\int_0^{2\pi}\int_0^{2\pi} \mathcal H_S \  dMdM_S+\frac{1}{4\pi^2}\int_0^{2\pi}\int_0^{2\pi} \mathcal H_M\  dMdM_M. 
\end{aligned}
\end{equation}
The Hamiltonian $\mathcal H_{sec}$ is the starting point to obtain long-term stability estimates for the secular geolunisolar model, either with normalization techniques, as in Section \ref{ssec:primo_art}, or through the application of stronger results, such as Nekhoroshev Theorem, as in the case of Section \ref{ssec:sec_art}. In order to carry on such investigation, one needs to express the above Hamiltonian in terms of \textit{action-angle variables}, such as the so-called \textit{modified Delaunay} ones (see \cite{morbidelli2002modern}), whose relation with the orbital elements is given by
\begin{equation}\label{eq:delaunay}
    \begin{cases}
        L=\sqrt{\mu_E a}\\
        P=\sqrt{\mu_Ea}\left(1-\sqrt{1-e^2}\right)\\
        Q=\sqrt{\mu_Ea}\sqrt{1-e^2}\left(1-\cos{i}\right)
    \end{cases}
    \begin{cases}
        \lambda=M+\omega+\Omega\\
        p=-\omega-\Omega\\
        q=-\Omega
    \end{cases}. 
\end{equation}
Note that, in terms of these new variables, the averaging performed above corresponds to the elimination of the fast angle $\lambda$, and, subsequently, to take the first action $L$ (and then the semimajor axis) as a constant, which we call $L_*=   \sqrt{\mu_E a_*}$. \\
As for the eccentricity and inclination, the presence of Sun and Moon on the ecliptic forces the existence of a circular, non-equatorial equilibrium orbit, whose inclination $i^{(eq)}$ depends on $a_*$ through the relation
\begin{equation*}
\begin{aligned}
  &i^{(eq)} \equiv i^{(eq)}(a_*) \simeq -\frac{A}{2B}\frac{1}{\left(\mu_E a_*\right)^{1/4}}, \\
  &A=-\frac{3 R_E^2 a_*^{7/4}\sin(2i_{0S})}{8\mu_E^{1/4}}\left(\frac{\mu_S}{a_S^3}+\frac{\mu_M}{a_M^3}\right)\\
  &B=\frac{3}{4}\frac{\mu_E^{1/2}R_E^2J_2}{a_*^{7/2}}+\frac{3\mu_S(2-3\sin^2(i_0S))}{16(\mu_E/a_*)^{1/2}a_S^3}+\frac{3\mu_M(2-3\sin^2(i_0M))}{16(\mu_E/a_*)^{1/2}a_M^3}. 
\end{aligned}
\end{equation*}\\
The equilibrium points $(e^{(eq)}, i^{(eq)})$, with $e^{(eq)}=0$, are traditionally called the \textit{forced elements} of the secular model, while the plane with inclination $i^{(eq)}$ is the \textit{Laplace plane}; more rigorous estimates on the value of $i^{(eq)}$ and its behaviour as a function of the distance can be found in \cite{rosengren2014classical}. \\
The stability estimates produced in this work are obtained by means of a \textit{semi-analytical} approach, namely, the application on rigorous analytical results on Hamiltonian computed numerically by means of the software \textit{Mathematica}\textsuperscript{\textcopyright}: for this reason, in Sections \ref{ssec:primo_art} and \ref{ssec:sec_art} we shall make use of a \textit{truncated} expression of $\mathcal H_{sec}$, whose truncation order will be specified case by case.

\subsection{Stability estimates through normal forms}\label{ssec:primo_art}

The first technique we propose to estimate the stability of the orbital elements in the secular geolunisolar model relies on  the application of a \textit{normal form} algorithm, and is similar to the one used in \cite{steichen1997long}. Before passing to the actual computation of the stability time in the satellites' case, a brief general introduction of the normal form theory is in order (a more complete dissertation on the subject can be found in \cite{DebCelEft2021satellites, efthymiopoulos2011canonical}). \\
Let us start by taking a Hamiltonian function expressed in action-angle variables
$\mathcal H(\underline J, \underline \theta)$, where $(\underline J, \underline \theta)\in U\times \mathbb T^n$, $n$ being the degrees of freedom of the system and $U\subset\mathbb R^n$ open. The principal aim of a normalization algorithm is to find a close-to-identity canonical transformation $\Phi:(\underline J, \underline \theta)\mapsto (\underline J', \underline\theta')$ such that the new Hamiltonian  $\mathcal H'=\mathcal H\circ \Phi^{-1}$ takes the form 
\begin{equation*}
    \mathcal H'(\underline J', \underline \theta')=Z(\underline J', \underline\theta')+R(\underline J', \underline\theta'),  
\end{equation*}
where: 
\begin{itemize}
    \item $Z(\underline J', \underline\theta')$ is the so-called \textit{normal part}, and has some desired property as, for example, the presence of first integrals of the motion; 
    \item $R(\underline J', \underline\theta')$ is the \textit{remainder}: in a suitable functional norm $\|\cdot\|$, it is such that $\|R\|\ll\|Z\|$.  
\end{itemize}
If the remainder's size is sufficiently small with respect to the normal part's ones, the overall dynamics under $\mathcal H'$ (and then under $\mathcal H$) can be considered as a small perturbation of the one induced by $Z$. As an example (which will be precisely our case), if $Z$ admits some integrals of the motion, such quantities are \textit{quasi-constant} for the whole $\mathcal H'$. The transformation $\Phi$ can be found by means of the \textit{Lie series} technique: its construction algorithm, which depends on the properties of the normal part we seek, is here omitted, and can be found  in \cite{efthymiopoulos2011canonical}.   \\
In this Section, a normalization algorithm is used to produce stability estimates for the eccentricity and inclination for orbits close to the equilibrium one, which has orbital parameters $(a_*, e^{(eq)}, i^{(eq)})$ (see Section \ref{sec:model}). 
As a preliminary step for this analysis, it is convenient to consider a set of Delaunay coordinates which are centered around the equilibrium, performing the change of coordinates 
\begin{equation}\label{eq:cambio star}
    \begin{cases}
        I_1=P^{(eq)}-P\quad & \phi_1=p\\
        I_2=Q^{(eq)}-Q\quad &\phi_2=q
    \end{cases}
\end{equation}
where $P^{(eq)}=\sqrt{\mu_E a_*}\left(1-\sqrt{1-(e^{(eq)})^2}\right)=0$ and $Q^{(eq)}\sqrt{\mu_Ea_*}\sqrt{1-(e^{(eq)})^2}\left(1-\cos{\left(i^{(eq)}\right)}\right)=\sqrt{\mu_Ea_*}\left(1-\cos{\left(i^{(eq)}\right)}\right)$. By means of a Taylor expansion, the Hamiltonian can be then written as a trigonometric polynomial in the square roots of the actions as 
\begin{equation}\label{eq:ham exp}
\begin{aligned}
    &\mathcal H(I_1,I_2, \phi_1, \phi_2)=&&\nu_1 I_1+\nu_2I_2\\
    &\quad&&+\sum_{s=3}^\infty\sum_{\substack{s_1,s_2\in\mathbb N \\ s_1+s_2=s}}\sum_{\substack{k_1,k_2\in\mathbb Z \\ |k_1|+|k_2|\leq s\\|k_1|+|k_2|\equiv s\  (mod\  2)}}h_{\substack{s_1,s_2\\k_1,k_2}}I_1^{s_1/2}I_2^{s_2/2}\cos(k_1\phi_1+k_2\phi_2). 
    \end{aligned}
\end{equation}
By construction and from Eq.\eqref{eq:delaunay}, one has that, for quasi-circular orbits close to the Laplace plane 
\begin{equation}\label{eq:order ie}
    I_1\simeq \sqrt{\mu_E a_*}\frac{e^2}{2}, \quad I_2\simeq \sqrt{\mu_E a_*} \frac{i^2}{2}: 
\end{equation}
where $(e,i)$ have to be intended as the differences with respect to the forced values $(0, i^{(eq)})$; this implies that, in the expansion \eqref{eq:ham exp}, the $s-$th term in the sum is of total order $s$ in eccentricity and inclination. \\
For computational reasons, in the following estimates the series in Eq.\eqref{eq:ham exp} is truncated up to order $N=15$. The rigorous procedure to obtain $\mathcal H(I_1, I_2, \phi_1, \phi_2)$ is described in \cite{DebCelEft2021satellites}, where one can also observe that the first order frequencies $\nu_1$ and $\nu_2$ are \textit{nearly equal}: this fact, which implies a $1:1$ resonance between the conjugate angles $\phi_1$ and $\phi_2$, will be crucial in the normalization procedure.  \\
Once the Hamiltonian is in the form of Eq. \eqref{eq:ham exp}, one can proceed with the normalization: in this case, it consists in finding a change of coordinates which makes the normal part depending only on the \textit{resonant angle} $\phi_1-\phi_2$, namely, a near-identity canonical transformation $\Phi$ such that the new Hamiltonian (which, with an abuse of notation, will be still called $\mathcal H(I_1,I_2, \phi_1,\phi_2)$) is given by the sum
\begin{equation}\label{eq:ham norm res}
    \mathcal H(I_1,I_2, \phi_1,\phi_2)=Z_{sec}(I_1, I_2)+Z_{res}(I_1, I_2, \phi_1-\phi_2)+R(I_1,I_2, \phi_1,\phi_2). 
\end{equation}
From a practical point of view, this result is achieved by applying to the initial Hamiltonian a sequence of transformations in the form of Lie series, aiming to remove the dependence on the angles, except for the resonant combination $\phi_1-\phi_2$, from all the terms of the series \eqref{eq:ham exp} up to order $M=12$. The choice of the normalization order $M$ is of particular importance to obtain optimal estimates: for more details on that, see \cite{fasso1989composition}. \\
Hamiltonians in the form of \eqref{eq:ham norm res} are usually said to be in \textit{resonant normal form}: here, the normal part is composed by the \textit{secular term} $Z_{sec}$, which does not depend on the angles, and the \textit{resonant} one, that depends on the actions as well as on the resonant combination $\phi_1-\phi_2$; as for the remainder, it is by construction of order $M$ in the square roots of the actions (namely, recalling Eq.\eqref{eq:order ie}, in eccentricity and inclination\footnote{We point out that, in principle, the variables $(I_1,I_2, \phi_1, \phi_2)$ are not the same as in \eqref{eq:ham exp}; nevertheless, since the new variables are obtained by means of \textit{near-identity} canonical transformations, they are the sum of the original ones and short period small variations, which do not affect their secular stability. }). \\ 
It is easy to prove that the quantity 
\begin{equation*}
I_1+I_2=\sqrt{\mu_Ea_*}\left[1-\sqrt{1-e^2}\left(1-\cos i\right)\right]\simeq \sqrt{\mu_Ea_*}\frac{e^2+i^2}{2}
\end{equation*}
is an integral of the motion for the dynamics induced by the sole normal part $Z_{sec}+Z_{res}$: the conservation of such quantity, which is equal to the vertical component of the satellite's angular momentum, determines a \textit{locking} between eccentricity and inclination, which can undergo only changes which keep constant the value of $I_1+I_2$. This fact, also known as \textit{Lidov-Kozai effect} (see \cite{Kozai, Lidov}), is common in many model of Celestial Mechanics which present resonance phenomena. \\
For the overall dynamics induced by \eqref{eq:ham norm res}, the quantity $I_1+I_2$ it is not constant anymore; nevertheless, if the remainder's norm is sufficiently small, is can be considered as  \textit{quasi-constant}, and it is possible to obtain stability estimates (namely, an upper bound for the time up to which it is bounded in a certain neighborhood around the initial values of $e$ and $i$) by measuring the size of $R$ in a suitable functional norm. \\
More precisely, let us fix a domain $\mathcal D\subset \mathbb R^2$ around the forced values for eccentricity and inclination $(0, i^{(eq)})$, and, given a function $f(e, i, \phi_1, \phi_2): \mathcal D\times \mathbb T^2\to \mathbb R$
let us consider the functional sup norm 
\begin{equation}\label{eq:sup norm}
    \|f\|_{\mathcal D, \infty}=\sup_{\substack{(e,i)\in \mathcal D\\(\phi_1,\phi_2)\in\mathbb T^2}}|f(e,i,\phi_1, \phi_2)|. 
\end{equation}
Our final objective is to evaluate the variation of $I_1+I_2$ along the trajectories induced by the normalized Hamiltonian in \eqref{eq:ham norm res}: to this aim, let us recall the relation 
\begin{equation*}
    \frac{d}{dt}(I_1+I_2)=\{I_1+I_2, \mathcal H\}=\{I_1+I_2, R\}, 
\end{equation*}
    where the notation $\{\cdot,\cdot\}$ denotes the Poisson brackets (see \cite{giorgilli2022notes}). Being $I_1+I_2$ a first integral for the normal part, $\{I_1+I_2, Z_{sec}+Z_{res}\}=0$, and then, given any $(\hat e, \hat i, \hat{\phi_1}, \hat{\phi_2})\in\mathcal D\times\mathbb T^2$, one has 
\begin{equation}\label{eq:derivata}
    \Bigg|\frac{d}{dt}(I_1+I_2)(\hat e, \hat i, \hat{\phi_1}, \hat{\phi_2})\Bigg |\leq \sup_{\substack{(e,i)\in\mathcal D\\(\phi_1,\phi_2)\in\mathbb T^2}}\Bigg|\frac{d}{dt}(I_1+I_2)( e,  i, {\phi_1}, {\phi_2})\Bigg | =\|\{I_1+I_2, R\}\|_{D, \infty}. 
\end{equation}
Let us now suppose that at the time $t=0$ the quantity $I_1+I_2$ has value $I_1^0+I_2^0$, corresponding to eccentricity and inclination $(e^0, i^0)\in \mathcal D$, and consider its time evolution over $t$. Suppose now to fix $\Gamma>0$ as the maximal variation allowed over a certain time for $(I_1+I_2)(t)$: applying the mean value theorem, it is possible to compute an upper bound for the time $T$ such that, for any $t\leq T$, 
\begin{equation*}
    |(I_1+I_2)(t)-(I_1^0-I_2^0)|\leq \Gamma. 
\end{equation*}
More precisely, from Eq.\eqref{eq:derivata} one has that 
\begin{equation*}
    |(I_1+I_2)(t)-(I_1^0-I_2^0)|\leq \|\{I_1+I_2, R\}\|_{D, \infty},  
\end{equation*}
so that 
\begin{equation*}
    T\geq \frac{\Gamma}{\|\{I_1+I_2, R\}\|_{D, \infty} }. 
\end{equation*}
The upper bound $\tilde T=\Gamma/\|\{I_1+I_2, R\}\|_{D, \infty}$ is the \textit{stability time} we seek: it depends of course on the maximal variation allowed $\Gamma$, as well as on the amplitude of the domain $\mathcal D$ in eccentricity and inclination we want to analyse. It is clear that the value of $\tilde T$ increases with $\Gamma$ and by taking smaller domains around the forced elements; moreover, it has a dependence on the reference value of the semimajor axis $a_*$. \\

For computational reasons, to produce the numerical estimates on $\tilde T$ the domain $\mathcal D$ is set to be $\mathcal D=\{(e,i)\in[0,0.1]\times [0\  rad, 0.1 \ rad]\}$, while $\Gamma$ will depend on $a_*$ through the relation
\begin{equation*}
    \Gamma=0.05\sqrt{\frac{\mu}{a_*}};  
\end{equation*}
additionally, the sup norm in \eqref{eq:sup norm} is replaced with an alternative functional norm based on majorization (see the details in \cite{DebCelEft2021satellites}). \\

It is clear that the whole stability estimate process depends crucially on the semimajor axis, which in the secular geolunisolar model is a constant parameter; for this reason, in the numerical estimates we distinguished within five different cases, which cover many different regimes (for the sake of clarity, they will be given in terms of the sum of the altitude with the Earth's radius): 
\begin{itemize}
    \item $a_*^{(1)}= 3\ 000\ km + R_E$, corresponding to an orbit just above the atmosphere; 
    \item $a_*^{(2)}=20\ 000\ km+R_E$, located in the MEO region; 
    \item $a_*^{(3)}=35\ 786\ km + R_E$, which corresponds to the altitude of GEO orbits; 
    \item $a_*^{(4)}=50\ 000 \ km + R_E$, corresponding to far object; 
    \item $a_*^{(5)}=100\ 000\ km + R_E$, which is the distance of objects very far from the Earth, where the influence of Sun and Moon is particularly strong. 
\end{itemize}
\begin{table}
\begin{tabular}{ |c|c|}
\hline
\textbf{Semimajor axis} & \textbf{Stability time $\boldsymbol{{\tilde T}}$}  \textbf{(years)}\\
\hline
$a_*^{(1)}$ & $4.61551\times 10^{13}$\\ 
\hline
 $a_*^{(2)}$ & $2.20144\times 10^{12}$\\  
 \hline
 $a_*^{(3)}$ & $3.51266\times 10^{10}$  \\
 \hline
 $a_*^{(4)}$ & $1.07263\times10^{8}$\\
 \hline
 $a_*^{(5)}$& $3.36609\times 10^{4}$  \\
 \hline
\end{tabular}
\caption{Stability time (in years) for the quantity $\mathcal I=\sqrt{\mu_E a_*}\left[1-\sqrt{1-e^2}(1-\cos{i})\right]$, obtained via normalisation method, for quasi circular and quasi equatorial orbits of the geolunisolar model and five different reference values of the semimajor axis. The data are taken from \cite{de2021semi}}. 
\label{tab:stab gls}
\end{table}

Table  \ref{tab:stab gls} shows the stability times obtained for these values of the semimajor axis; as one can easily notice, though particularly long, the time $\tilde T$ decreases with the altitude, with a significant worsening beyond GEO distance. \\
These results, obtained numerically, are  consistent with the theory: for small values of the semimajor axis the secular geolunisolar model can be well approximated by the secular $J_2$ model (namely, the model in which only the geopotential up to the term $J_2$ is considered, averaged over the mean anomaly), which is integrable; on the other hand, going farther from the Earth's surface, the influence of Sun and Moon gets stronger and stronger, leading to a perturbation that produces instability in  the model. 

\subsection{Exponential stability estimates through Nekhoroshev Theorem}\label{ssec:sec_art}

\begin{figure}
    \centering
    \includegraphics{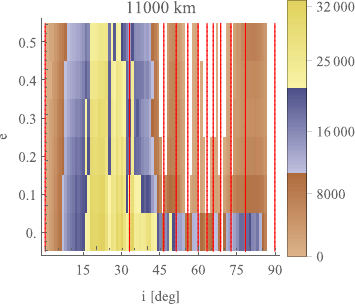}
    \includegraphics{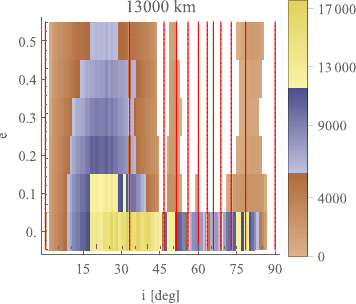}
    \includegraphics{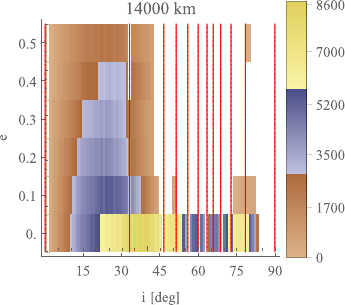}
    \includegraphics{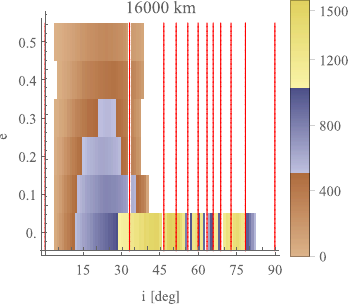}
    \includegraphics{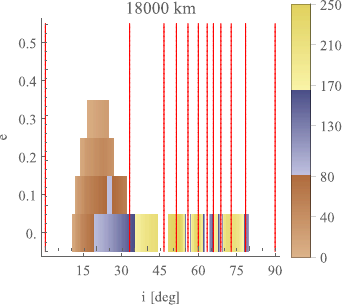}
    \includegraphics{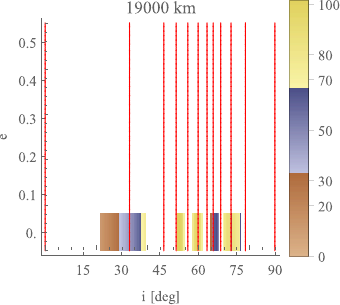}
    \caption{Stability times computed for different values of semimajor axis, eccentricity and inclination using the nonresonant version of Nekhoroshev theorem. The color scale refers to the computed stability times (in years), while the white region correspond to the values of $(e,i)$ where Theorem \ref{thm:nekh} can not be applied with the present algorithm. The red lines are in correspondence of the inclinations of the secular geolunisolar resonances. Data taken from \cite{cellettiDeBlasiEft2023nekhoroshev}. }
    \label{fig:nekh stab time} 
\end{figure}

In addition to be used as in Section \ref{ssec:primo_art} to produce stability estimates based on the mean value theorem, normal form algorithms are an essential preliminary tool (as well as a proving strategy) to apply the celebrated \textit{Nekhoroshev theorem} (see \cite{nekhoroshev1977exponential, Poschel}), here presented in its nonresonant form. In general (we will be more precise in Theorem \ref{thm:nekh} for the specific case of the nonresonant regime), such theorem can be applied to \textit{quasi-integrable} Hamiltonians of the form 
\begin{equation}\label{eq:nekh ham}
    \mathcal H(\underline J, \underline \theta)=h_0(\underline J)+h_1(\underline J, \underline \theta), 
\end{equation}
where $h_0$ depends only on the actions and $h_1$ depends on the angles as well. As a consequence of Hamilton's equations, the dynamics induced by $h_0$ has the actions as first integrals of the motion, namely, $\underline J(t)=\underline J_0$ for any $t\geq0$, $\underline J_0$ being their initial value. Under some suitable nondegeneracy condition on $h_0$ and provided that the perturbative function $h_1$ is small enough, it is possible to estimate the stability time of the actions under the dynamics induced by the whole $\mathcal H$: in particular, it is possible to find an open set around $J_0$ where the actions are bounded for a time which is \textit{exponentially long} in the inverse of the perturbation's norm. \\
In the most general formulation of Nekhoroshev theorem, the nondegeneracy hypothesis required on $h_0$, called \textit{steepness} condition, resumes essentially in asking for a quantitative transversality condition for the gradient $\nabla h_0$; here, we will rely on a simpler \textit{nonresonance} hypothesis, based on the non-commensurability of the coefficients of the actions at first order, which can be easily verified numerically. As we will see while presenting the numerical results (see Figure \ref{fig:nekh stab time}), the application of this simpler version of the theorem implies a cost in terms of the region of the $(a,e,i)-$space where our estimates hold: nevertheless, the stability times obtained are particularly good in a strip of the  MEO region and in a nonresonant regime, being comparable with the satellites' average orbital lifetime; a finer analysis, considering the geometry of the resonances in the geolunisolar problem, is anyway possible. \\
To apply the nonresonant version of Nekhoroshev theorem to our geolunisolar case, it is necessary to put the Hamiltonian \eqref{eq:ham sec} in the form of a sum of an integrable term and a perturbation, as in Eq. \eqref{eq:nekh ham}: to this aim, we will rely again on a normal form algorithm. \\

\paragraph{\textbf{Hamiltonian preparation}}\label{ssec:normalizzazione 2}
Let us start again from the secular geolunisolar Hamiltonian as presented in Eq.\eqref{eq:ham sec}. While in  Section \ref{ssec:primo_art} we focused our investigation in a small neighborhood (in eccentricity and inclination) of the forced elements $(0, i^{(eq)})$, here we aim to provide stability times holding for values of the orbital parameters which are not necessarily small. We will then produce a \textit{sequence} of Hamiltonian functions, each of which is obtained by expanding $\mathcal H_{sec}$ around a triplet of reference values $(a_*,e_*,i_*)$ in a grid covering the set $[11\ 000\  km, 20\ 000 \ km]\times[0,0.5]\times[0^\circ,90^\circ]$; for each Hamiltonian of such sequence, we will follow a numerical procedure, described below, to provide stability estimates holding in a neighborhood of the corresponding reference values $(e_*, i_*)$ (remember that, in the secular geolunisolar problem, the semimajor axis $a_*$ is \textit{a priori} constant for any forward time). \\
In practice, once fixed $(a_*, e_*, i_*)$, one can perform a translation in the actions analogous to the one presented in Eq. \eqref{eq:cambio star} to obtain an expansion of the form 
\begin{equation}\label{eq:ham exp 2}
    \mathcal H(I_1,I_2,\phi_1,\phi_2)=\sum_{j=1}^\infty g^{(j)}(I_1,I_2, \phi_1,\phi_2), 
\end{equation}
where the $j-$th term of the sum is given by 
\begin{equation}
\begin{aligned}
    &g^{(1)}(I_1, I_2, \phi_1,\phi_2)=\omega_1 I_1+\omega_2 I_2\\
    &g^{(j)}(I_1,I_2, \phi_1,\phi_2)=\sum_{\substack{\underline l=(l_1,l_2)\in\mathbb Z^2\\l_1+l_2=j}}a_{l_1,l_2}^{(j)}I_1^{l_1}I_2^{l_2}+\sum_{\substack{\underline l=(l_1,l_2)\in \mathbb Z^2\\\underline k=(k_1,k_2)\in \mathbb Z^2\\ l_1+l_2=j-2}} b_{\substack{l_1,l_2\\k_1,k_2}}^{(j)}I_1^{l_1}I_2^{l_2}e^{i\left(k_1 \phi_1+k_2\phi_2\right)}, \quad j\geq 2. 
\end{aligned}
\end{equation}
Note that the expansion in Eq. \eqref{eq:ham exp 2} is the analogous of Eq.\eqref{eq:ham exp} in Section \ref{ssec:primo_art}, although in this case the exponential form has been chosen. The explicit expressions of $\omega_1$ and $\omega_2$, as well as of the coefficient of $a^{(j)}$ and $b^{(j)}$ at first and second order, can be found in \cite{cellettiDeBlasiEft2023nekhoroshev}. By computing $\omega_1$ and $\omega_2$ numerically, it is possible to observe that there are particular values of the reference inclination $i_*$ for which they are \textit{commensurable}: these are the inclinations of the so-called \textit{secular resonances} for the geolunisolar problem (see for example \cite{breiter1999lunisolar}), which will play a fundamental role in the upcoming stability analysis. \\
As in Section \ref{ssec:primo_art}, for computational reason the sum in Eq.\eqref{eq:ham exp 2} has been truncated, up to order $N=12$. \\
The next step towards producing stability estimates via Nekhoroshev theorem consists in normalising the Hamiltonian in Eq. \eqref{eq:ham exp 2}, namely, using canonical transformations to obtain an expression as in Eq.\eqref{eq:nekh ham}, where $h_0$ contains only angle-independent terms and the size of $h_1$ can be controlled with a suitable norm. \\
Let us start by considering the non-normalised sum in Eq. \eqref{eq:ham exp 2}, and suppose to split it into the form 
\begin{equation}
\begin{aligned}
\mathcal H(I_1, I_2, \phi_1,\phi_2)=\tilde h_0(I_1,I_2)+\tilde h_1(I_1,I_2,\phi_1, \phi_2), 
\end{aligned}
\end{equation}
where $\tilde h_0$ contains only the angle-independent terms in \eqref{eq:ham exp 2} and $\tilde h_1$ contains all the others. If we suppose that the action values are bounded, is clear that, from $j=4$ on, the size of the angle-dependent summands decrease quadratically with the action's bound; on the other hand, the purely trigonometric terms, as well as angle-dependent ones which are linear in the actions, are harder to control\footnote{It  can be showed that the presence of purely trigonometric terms is related to the presence of the Laplace plane, see for example the explicit expressions of the coefficients in \cite{cellettiDeBlasiEft2023nekhoroshev}}. The normalisation algorithm performed in this case aims precisely to the elimination of such terms up to a certain order $M$ (in the actual computation, $M$ is set equal to $6$) via a sequence of suitable Lie series transformations. The complete algorithm, whose extended description can be found in \cite{cellettiDeBlasiEft2023nekhoroshev}, leads finally to a new Hamiltonian 
\begin{equation}\label{eq:ham norm 2}
    \mathcal H_{norm}(I_1,I_2, \phi_1,\phi_2)= \tilde \omega_1 I_1+\tilde \omega_2I_2+Z(I_1,I_2,\phi_1,\phi_2)+R(I_1, I_2,\phi_1,\phi_2), 
\end{equation}
where, with an abuse of notation, the new action-angle variables are still called $I_i$ and $\phi_i$, $i=1,2$. In Eq.\eqref{eq:ham norm 2}, the normal part, composed by the linear part plus $Z$, is composed by angle-independent terms plus other terms which could be angle-dependent but at least quadratic in the actions. As for the remainder term $R$, it could contain terms which depend on $(\phi_1,\phi_2)$ and are constant or linear in the actions; nevertheless, provided the normalization algorithm converges (namely, the size of the coefficients of the remainder decreases in the process), such terms are \textit{small} with respect to $Z$. \\
The convergence of the normalization is a crucial issue of the overall procedure, which depends heavily on the non-commensurability of the initial frequencies $\omega_1$ and $\omega_2$; furthermore, such convergence influences also the final value of the frequencies, denoted by $\tilde \omega_1$ and $\tilde \omega_2$. The variation in such quantities is negligible whenever the normalisation converges. \\
The normalized Hamiltonian $\mathcal H_{norm}$ is the starting point to obtain exponential stability estimates via non-resonant Nekhoroshev theorem, which we now recall in the version by P\"oschel (see \cite{Poschel}), after some useful definitions. \\

Let us start by considering, in general, a Hamiltonian of the form 
    \begin{equation}
        \mathcal H(\underline J,\underline \theta)=h_0(\underline J)+h_1(\underline J,\underline \theta), 
    \end{equation}
    which is assumed to be real analytic in $(\underline J,\underline \theta)\in A\times \mathbb T^n$, $A\subset\mathbb R^n$. Suppose also that the above Hamiltonian can be extended analytically to the set 
    \begin{equation}\label{eq:dominio}
        \begin{aligned}
            &D_{r_0,s_0}=A_{r_0}\times \mathbb S_{s_0}\\
            &A_{r_0}=\left\{\underline J\in \mathbb C^n \ :\ dist(\underline J,A)<r_0\right\}\\
            &S_{r_0}=\left\{\underline \theta\in\mathbb C^n\ :\ Re(\theta_j)\in\mathbb T, \ \max_{j=1, \dots, n}|Im(\theta_j)|<s_0\right\},
        \end{aligned}
    \end{equation}
where $r_0$ and $s_0$ are two positive real constants. As a last assumption, let us suppose that the Hessian matrix associated to $h_0$ is bounded in $A_{r_0}$, namely, that there exists a constant $\mathcal M>0$ such that, denoted with $\|\cdot\|_o$ the operator norm induced by the Euclidean one on $\mathbb R^2$,  
\begin{equation}
    \sup_{\underline J\in A_{r_0}}\|\nabla^2 h(\underline J)\|_o\leq \mathcal M. 
\end{equation}
Given now an analytic function expressed as 
\begin{equation}
    g(\underline J,\underline \theta)=\sum_{\underline k\in \mathbb Z^n} g_{\underline k}(\underline J)e^{i\underline k\cdot \underline \theta}, 
\end{equation}
we define the \textit{Cauchy norm} of $g$
\begin{equation}
    |g|_{A,r_0,s_0}=\sup_{\underline J\in A_{r_0}}\sum_{\underline k\in\mathbb Z^n}|g_{ \underline k}(\underline J)|e^{|\underline k|s_0}, 
\end{equation}
where $|\underline k|=|k_1|+\dots+|k_n|$. \\
\begin{theorem}\label{thm:nekh}
Given $\alpha, K>0$ suppose that $D\subseteq A$ us a completely $\alpha-K-$nonresonant domain, namely, 
\begin{equation}
\forall \underline k\in\mathbb Z^n\setminus \{\underline 0\}, \ |\underline k|\leq K, \ \text{and}\ \forall \underline J\in D \ \text{one has }|\underline k\cdot \nabla h_0(\underline J)| \geq   \alpha.     
\end{equation}
Let $a,b>0$ such that $a^{-1}+b^{-1}=1$; if 
\begin{equation}\label{eq:nekh hyp}
    |h_1|_{A, r_0, s_0}\leq \frac{1}{2^7}\frac{\alpha r}{K}=\epsilon^*, \quad r\leq\min\left(\frac{\alpha}{a \mathcal M K}, r_0\right)
\end{equation}
then for every orbit of initial conditions $(\underline J_0, \underline \theta_0)\in D\times \mathbb T^n$ one has 
\begin{equation}\label{eq:tstab}
    \|\underline J(t)-\underline J_0\|\leq r\ for\ |t|\leq \frac{s_0 r}{5 |h_1|_{A,r_0, s_0}}e^{Ks_0/6}=T_{stab}. 
\end{equation}
\end{theorem}
Once one has precise numerically computed values for all the quantities involved, one can use Theorem \ref{thm:nekh} to produce stability estimates for the actions; more precisely, in a non-resonant regime defined through the notion of $\alpha-K-$nonresonance, one can find an open set in $\R^n$ in which the actions are bounded for a time which is exponentially long in $K$. We stress that such result is to be intended as \textit{local}, in the sense that it holds for initial values for the actions in a subset $D$ of $A$. Numerical evidences (see the paragraph "Numerical results" below) show how the cut-off value $K$ satisfies a relation of the type $K\sim \left(c_1 |h_1|_{A,r_0,s_0}\right)^{-c_2}$, $c_1$ and $c_2$ being two positive constants. Such behaviour is consistent with theoretical results (see for example \cite{Poschel}), and allows to conclude that the stability time is \textit{exponentially long} with respect to the perturbation's norm to some power. \\
\paragraph{\textbf{Numerical results}} To produce stability times though Theorem \ref{thm:nekh} for the secular geolunisolar model, it is necessary to set up an algorithm that, given reference values of the orbital elements, computes the quantities involved in the Theorem and finally, if the hypothesis \eqref{eq:nekh hyp} is satisfied, provides $T_{stab}$ as in Eq.\eqref{eq:tstab}. In practice, our algorithm develops into the below steps.
\begin{enumerate}
    \item We start by fixing the constants $a,b,r_0,s_0$, whose value has been established by trials and errors, and could be possibly tuned to obtain optimal estimates. In particular, we impose $r_0=s_0=0.1$, $a=9/8$ and $b=1/8$ (the choice of $a$ and $b$'s values is the same one can find in \cite{Poschel}). Moreover, we fix a reference value of the semimajor axis $a_*$, which, by virtue of the averaging process, is constant along every orbit. 
    \item Fixed the values $(e_*,i_*)$, we compute numerically the expansion \eqref{eq:ham exp 2}, to arrive, after the normalization, to the form \eqref{eq:ham norm 2}. We stress that the final values of the frequencies $\tilde\omega_1, \tilde\omega_2$, as well as the actual size of the remainder term $R$, depend heavily on the non-commensurability of $\omega_1$ and $\omega_2$, namely, on the reference values $(e_*, i_*)$. This means that different values of eccentricity and inclination could lead to completely different outcomes in terms of normalization. The Hamiltonian can be now splitted into an integrable part $h_0(I_1,I_2)$ containing only the angle-independent terms plus a perturbation $h_1(I_1,I_2,\phi_1,\phi_2)$ which contains all the other terms. 
    \item We can now compute the quantities involved in Theorem \ref{thm:nekh}: first of all, we define the actions' set $A$ as 
    \begin{equation}
        A=[I_1^*-0.1,I_1^*+0.1]\times [I_2^*-0.1,I_2^*+0.1],  
    \end{equation}
    $I_1^*, I_2^*$ being the actions corresponding to the reference values; one can then define $D_{r_0,s_0}$ as in Eq. \eqref{eq:dominio} and 
    \begin{equation}
        \mathcal M=\sup_{(I_1,I_2)\in A_{r_0}} \|\nabla^2 h_1(I_1,I_2)\|_o.
    \end{equation} 
    As for the nonresonance parameters $\alpha, K$, we search for their optimal values, provided condition \eqref{eq:nekh hyp} is satisfied, as follows: for every $i=1, \dots, 50$ we compute 
    \begin{equation}
        \alpha_i=\min_{\substack{\underline l=(l_1,l_2)\in\mathbb Z^2\\|\underline l|\leq i}}\{\omega_1 l_1+\omega_2 l_2\}, \ r_i=\min\left\{\frac{\alpha_i}{a \mathcal M i}\, r_0\right\}, \ \epsilon^*_i=\frac{1}{2^7b}\frac{\alpha_ir_i}{i}. 
    \end{equation}
    At this point, one can compute $|h_1|_{A,r_0,s_0}$ and check whether there exists $i\in\{1, \dots,50\}$ such that $|h_1|_{A,r_0,s_0}\leq \epsilon^*_i$: if it happens, then one can take $K$ as the maximal $i$ such that the condition is verified, $\alpha=\alpha_K$ and compute the stability time as in Eq.\eqref{eq:tstab}. On the other hand, since the sequence $\{\epsilon^*_i\}_{i=1}^{50}$ is clearly decreasing, if $|h_1|_{A,r_0,s_0}>\epsilon_1^*$ there is no hope for the theorem to be applied for the specific values $(a_*,e_*,i_*)$ and any $i\in\{1, \dots, 50\}$: in this case, we impose $K=0$. 
\end{enumerate}

Figure \ref{fig:nekh stab time} shows the numerical results obtained for  semimajor axis' values from $11\ 000\ km$ and $19\ 000\ km$ and $(e,i)$ ranging into a mesh of $[0,0.5]\times[0^\circ,90^\circ]$ of step $0.1$ in eccentricity and $0.5^\circ$ in inclination. The color scale indicates the value of the stability time (in years) obtained, while the white region of such values of $(e_*,i_*)$ for which condition \eqref{eq:nekh hyp}, using the proposed algorithm, does not hold. The red lines are put in correspondence of the known \textit{inclination-dependent} resonances for the secular geolunisolar problem (see \cite{breiter1999lunisolar}). \\
From the numerical results, it is evident as the domain where the Theorem can be applied shrinks manifestly with $a_*$, and, concurrently, the estimates on the stability times get worse. Moreover, an evident influence of the resonances comes out, since, even in the best case (i.e. for $a_*=11\ 000\ km$), white regions around the corresponding inclinations appear. \\
The role played by the resonances in the overall procedure enters at two different levels: during the normalization process and, later, in the very application of Theorem \ref{thm:nekh}. \\
As for the first normalization, one can check from the explicit expression of the coordinate changes used to remove the ``unwanted" terms from the normal part (see \cite{cellettiDeBlasiEft2023nekhoroshev} for all the details) that linear combinations of $\omega_1$ and $\omega_2$ appear at the denominator: whenever the frequencies are resonant, such denominators (the so called \textit{small divisors}) approach zero, leading to an explosion in the remainder $R$ and, subsequently, in the size of $h_1$. We refer again to \cite{cellettiDeBlasiEft2023nekhoroshev} for a detailed analysis of the convergence of the first normalization, including results on the change in the frequencies' value during the process.\\ 
On the other hand, the simple fact that we are using a \textit{nonresonant} version of Nekhoroshev theorem makes clear how having a commensurability relation at low order between $\tilde \omega_1$ and $\tilde \omega_2$ correspond to a value of $\alpha$ (and, as a consequence, of the threshold $\epsilon^*$) drastically low, making nearly impossible for the norm of $h_1$ to remain below $\epsilon^*$. \\
The effect of the distance on the worsening of the results has a more complex reason, which can be explained, roughly, by the following heuristic argument: it can be shown that, after the normalization, the remainder $R$ contains purely trigonometric terms whose size is comparable to the one of $\left(C a_*^5 \tan{i^*}\right)^M$, where $C$ is a suitable constant and $M$ is the normalization order, here put equal to $6$. As a consequence, the size of these terms, which can not be controlled by taking a smaller domain in the actions, grows swiftly with $a_*$ and whenever $i_*$ approaches $90^\circ$. \\
We conclude the exploration of the numerical analysis of the stability problem by providing an example which shows the behaviour of the computed value of the cut-off value $K$ with respect to the perturbation's norm $|h_1|_{A,r_0,s_0}$. 
\begin{figure}
    \centering
    \includegraphics{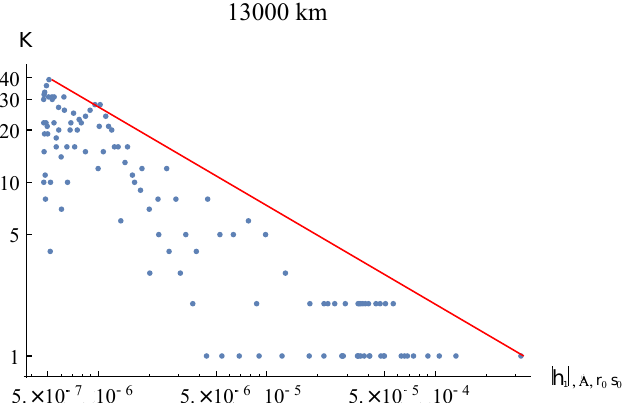}
    \caption{$LogLog$ plot of the points $\{|h_1|_{}A,r_0,s_0, K\}$ for $a_*=13\ 000\ km$, $e_*=0.2$ and $i_*\in[0^\circ,90^\circ]$. Data taken from \cite{cellettiDeBlasiEft2023nekhoroshev}.  }
    \label{fig:cut off}
\end{figure}
Figure \ref{fig:cut off} shows the LogLog plot of the values of $K$ and $|h_1|_{A,r_0,s_0}$ for $a_*=13\ 000$ km, $e_*=0.2$ and inclinations in a mesh of $[0^\circ, 90^\circ]$. It is evident a relation of the type 
\begin{equation}
    K\leq (c_1 |h_1|_{A,r_0,s_0})^{-c_2}, 
\end{equation}
which is consistent with the expected theoretical results, and allows to conclude, as anticipated before, that the final estimates can be actually considered as exponentially long in the inverse of the perturbing function's norm.

\subsection{Further considerations and conclusions}\label{ssec:conclusioni sat}
The techniques used in Sections \ref{ssec:primo_art} and \ref{ssec:sec_art} are examples of how semi-analytical manipulations in a Hamiltonian framework could be used to gain information in the long-term dynamics of a body orbiting around the Earth under the influence of the latter, Sun and Moon. Other examples of this kind can be found in the literature (see for example \cite{DebCelEft2021satellites}, where also the case of the $J_2-$model is taken into consideration, or \cite{KingHele, celletti2017analytical, Rosengren2013}). The first method, inspired by the work of Giorgilli and Steichen in \cite{steichen1997long}, essentially provides stability times which, though very long (see Table \ref{tab:stab gls}), are linear with respect to the perturbation's norm, and hold in quasi-circular orbits lying close to the Laplace plane. As for the second method, it produces estimates which are \textit{exponentially long} with respect to the inverse of the perturbation's size, showing all the potential of the Nekhoroshev theorem (see \cite{nekhoroshev1977exponential}), which could be used also in higher dimensions; on the other hand, at present the domain in which the results are truly substantial is not particularly large (see Figure \ref{fig:nekh stab time}). Nevertheless, we stress that different strategies to obtain an initial normal form may overcome the convergence problem, and, most of all, that the above procedure is based on a \textit{nonresonant} result: a finer analysis of the geometry of the resonances in the secular geolunisolar problem would allow to use Nekhoroshev theorem in its complete version, obtaining estimates valid in a resonant regime as well. \\
As for the model we chose to use, we stress that we are considering the influence of the geopotential only up the $J_2-$term. The overall analysis can be refined by taking also further terms in \eqref{eq:ham J2}, like for example the ones corresponding to $J_2^2$, $J_3$ and $J_4$. A comparison between our results and the ones one can obtain by considering this more complete model is presented at the end of \cite{cellettiDeBlasiEft2023nekhoroshev}, showing that, in the practical context on the satellites' motion, the stability times obtained for the two models, though different, are so long with respect to the average operational lifetime that any change does not really affect the validity of the estimates.

\section{Regular and chaotic motions in Galactic Billiards}\label{sec:biliardi}

The current Section resumes the results contained in \cite{deblasiterracinirefraction, deblasiterraciniOnSome,barutello2023chaotic} on the analysis of the refraction galactic billiard (see Section \ref{sec:intro}), a model aiming to provide a simplified description of the motion of a particle in an ellipsoidal galaxy having a central super-massive core. \\
Section \ref{ssec:bil modello} is intended as a description, in this framework, of the considered dynamical system, complete with the motivations that led us to choose a the considered potentials, as well as a refractive interface. 
The results obtained are divided into three subgroups: Section \ref{ssec:bil primo_art} is focused on the existence and linear stability of equilibrium trajectories for the model, and provides numerical and analytical evidences of bifurcation phenomena regarding a particular class of orbits. Section \ref{ssec:bil secondo_art} extends the analysis to periodic and quasi-periodic trajectories, within a perturbative regime constructed by taking into consideration quasi-circular billiards. In Section \ref{ssec:bil terzo_art}, the problem of the possible arising of chaotic behaviours is taken into account, arriving to the detection of simple geometric conditions on the billiard's boundary that ensure the presence of a chaotic subsystem at high energies.

\subsection{The model: analytical set-up and motivations}\label{ssec:bil modello}

Let us start the analytical description  of our galactic refraction billiard by taking a smooth open domain $D \subset \R^2$, containing the origin, and considering the potential
\begin{equation}
    V(z)=
    \begin{cases}
        V_I(z)=\mathcal E+h +\frac{\mu}{|z|} &\text{if } z\in D\\
        V_E(z)=\mathcal E-\frac{\omega^2}{2}|z|^2 \quad &\text{if }z\notin D 
    \end{cases}
\end{equation}
where $\mathcal E, \omega, h, \mu$ are positive constants representing respectively the energy and frequency of the outer harmonic oscillator, the difference in energy between inner and outer trajectories and the central body's mass parameter. Starting from initial conditions on the interface $\partial D$, the trajectories at zero energy induced by the inner potential are Keplerian hyperbol\ae, while the outer ones are elliptic harmonic arcs: with a \textit{broken geodesics technique} (see \cite{Seif}) we can construct complete trajectories in our system by patching together outer and inner arcs. The connection rule is given by the refraction Snell's law described in Figure \ref{fig:snell}, left: 
\begin{figure}
    \centering
    \begin{overpic}[width=.3\linewidth]{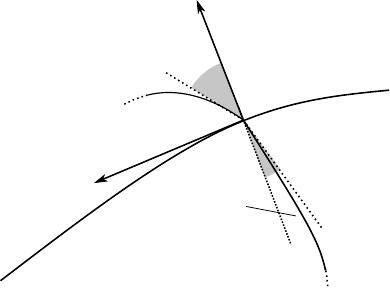}
    \put (45,59) {\rotatebox{40}{\tiny$\alpha_E$}}
    \put (55,20) {\rotatebox{20}{\tiny$\alpha_I$}}
    \put (65,42) {\rotatebox{0}{\tiny$z$}}
    \end{overpic}
    \quad
    \begin{overpic}[width=.3\linewidth]{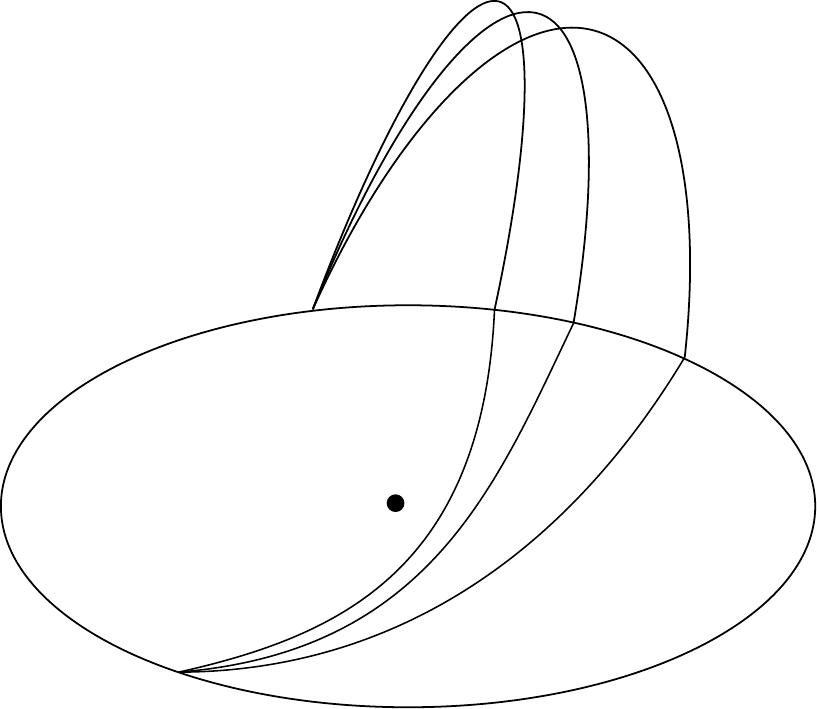}
    	\put (35,42) {\rotatebox{0}{\tiny$p_0$}}
    	\put (20,7) {\rotatebox{0}{\tiny$p_1$}}
    	\put (55,44) {\rotatebox{0}{\tiny$p$}}
    \end{overpic}
    \caption{Left: Snell's refraction law. The angles $\alpha_E$ and $\alpha_I$ are the angles respectively of the outer and the inner arc with respect to the normal direction to $\partial D$ in $z$. The two angles are connected by the relation \eqref{eq:snell}. Right: concatenations from $p_0$ to $p_1$ with an outer and inner arc, for different positions of the transition point $p$. The left figure is taken from \cite{barutello2023chaotic}.  }
    \label{fig:snell}
\end{figure}
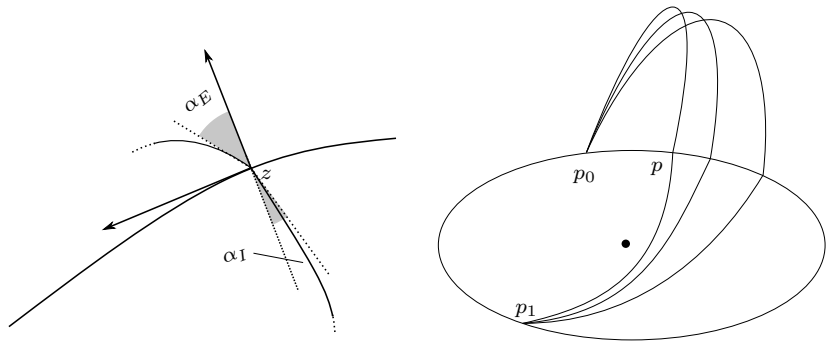
denoting with $\alpha_I$ and $\alpha_E$ respectively the angles of the inner and outer arcs connected at a point $z\in\partial D$ with respect to the normal direction to $\partial D$ in $z$, the following relation must be satisfied
\begin{equation}\label{eq:snell}
    \sqrt{V_I(z)}\sin\alpha_I=\sqrt{V_E(z)}\sin\alpha_E. 
\end{equation}
Geometrically, Eq.\eqref{eq:snell} translates in the conservation of the tangent component of the velocity after the transition. \\
The choice of this kind of connection rule is based on different arguments: first of all, from a physical point of view, it can be seen as a generalisation for non-constant potentials and non-straight interfaces of the classical Snell's law for light rays.
On the other hand, it has a rigorous and robust variational interpretation, which will be crucial in the whole forthcoming analysis. To explain it (see \cite{deblasiterracinirefraction} for further details), let us consider a concatenation of an outer and a inner arc that connects two point on the boundary $p_0$ and $p_1$, passing trough a transition point $p$ (see Figure \ref{fig:snell}, right). It is possible to associate to any of the two arcs, denoted for the moment by $z_E(t)$ and $z_I(t)$, the corresponding \textit{Jacobi lengths}
\begin{equation}\label{eq:Jacobi dist}
    \mathcal L_{E/I}(z_{E/I})=\int_0^{T_{E/I}}|z'_{E/I}(t)|\sqrt{V_{E/I}\left(z_{E/I}(t)\right)} dt, 
\end{equation}
where $z_E(0)=p_1$, $z_E(T_E)=z_I(0)=p$ and $z_I(T_I)=p_1$. Under suitable conditions, it can be proved that the outer (resp. inner) trajectory under the potential $V_E$ (resp. $V_I$) arc connecting two points on the boundary is unique: as a consequence, the functions $\mathcal L_E$ and $\mathcal L_I$ depend only on the endpoints. 
The inner and outer Jacobi lengths can be combined to obtain the \textit{total Jacobi length} of our concatenation, and, making use of this quantity, it is possible to state Snell's law in a variational way as follows: we say that the concatenation from $p_0$ to $p_1$ through $p$ satisfies Snell's law at the transition point if and only if $p$ is a critical point for the total Jacobi length of the concatenation itself, that is,
\begin{equation}\label{eq:snell var}
    \nabla_{\tilde p}\left(\mathcal L_E(p_0,\tilde p)+\mathcal L_I(\tilde p, p_1)\right)_{\tilde p=p}=
    \begin{pmatrix}
        0\\0
    \end{pmatrix}. 
\end{equation}
Of course, an analogous reasoning applies whenever the transition is from inside to outside. \\

As customary in billiards theory, to study the two dimensional dynamics of the trajectories of the complete system it is possible to restrict ourselves to a discrete map which keeps track of the behaviour of a concatenation whenever it hits the boundary: this is the so-called \textit{first return map}, which, starting from generic initial conditions on the boundary (position and velocity vector), summarises the behaviour of the generated trajectory after every concatenation of an outer and subsequent inner arc. \\
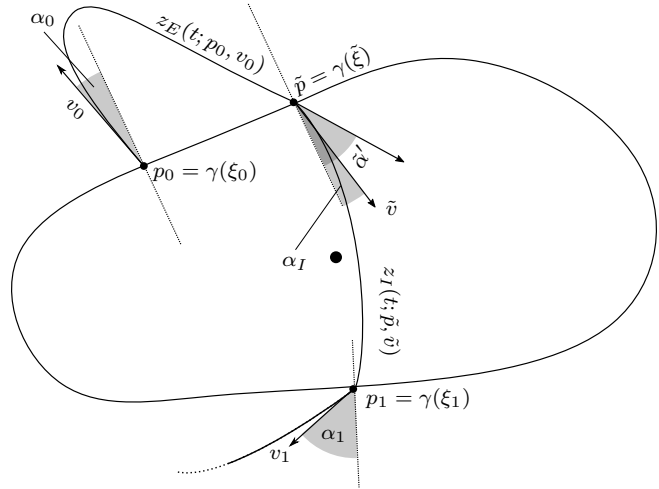
\begin{figure}
    \centering
    \begin{overpic}[width=.5\linewidth]{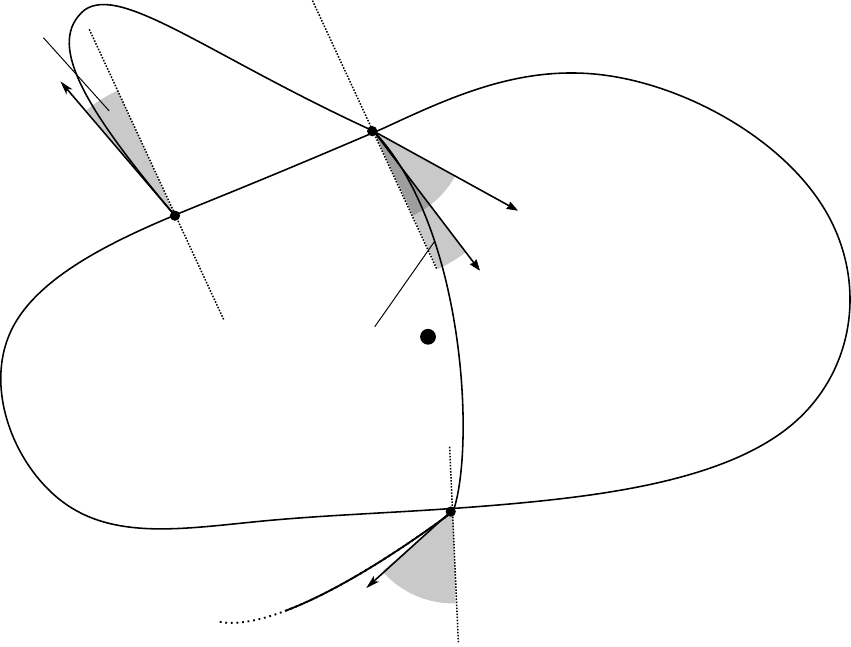}
    	\put (22,49) {\rotatebox{0}{\tiny$p_0=\gamma(\xi_0)$}}
    	\put (8,60) {\rotatebox{-35}{\tiny$v_0$}}
    	\put (3,73) {\rotatebox{0}{\tiny$\alpha_0$}}
    	\put (22,73) {\rotatebox{-25}{\tiny$z_E(t; p_0,v_0)$}}
    	\put (43,62) {\rotatebox{25}{\tiny$\tilde p=\gamma(\tilde\xi)$}}
    	\put (42,35) {\rotatebox{0}{\tiny$\alpha_I$}}
    	\put (52,50) {\rotatebox{60}{\tiny$\tilde\alpha'$}}
    	\put (58,43) {\rotatebox{0}{\tiny$\tilde v$}}
    	\put (57,35) {\rotatebox{-90}{\tiny$z_I(t; \tilde p , \tilde v)$}}
    	\put (55,13.5) {\rotatebox{0}{\tiny$p_1=\gamma(\xi_1)$}}
    	\put (48,8) {\rotatebox{0}{\tiny$\alpha_1$}}
    	\put (40,5) {\rotatebox{0}{\tiny$v_1$}}
    \end{overpic}
    \caption{First return map: starting from initial conditions $(p_0, v_0)$, determined by the one-dimensional parameters $(\xi_0,\alpha_0)$, the trajectory is follow through an outer arc, a refraction from outside to inside, an inner arc and a refraction from inside to outside to find the final conditions $(p_1,v_1)$, defined by $(\xi_1,\alpha_1)$. }
    \label{fig:first return}
\end{figure}
To be more precise, let us start by parametrising $\partial D$ with a smooth, closed and simple curve $\gamma:I\to \R^2$, $\xi\mapsto \gamma(\xi)$, where $I\subset \R$ is a suitable interval. For the sake of simplicity and without loss of generality, we can suppose that $\gamma$ is the arc length parametrisation of $\partial D$, so that $|\dot \gamma(\xi)|=1$ for any $\xi\in I$. Let us now take initial conditions on the boundary for an outer arc, $(p_0,v_0)\in \partial D\times \R^2$, such that $v_0$ points outside $D$ and the energy conservation law for the outer problem is satisfied, that is, $|v_0|^2/2-\mathcal E+\omega^2|p_0|^2/2=0$ (see Figure \ref{fig:first return}). Such initial conditions are uniquely determined by a pair of one dimensional parameters $(\xi_0,\alpha_0)$, with $\gamma(\xi_0)=p_0$ and $\alpha_0$ the angle between $v_0$ and the outward-pointing normal unit vector to $\gamma$ in $\xi_0$. Once the initial conditions are fixed, we can consider the outer arc $z_E(\cdot; p_0, v_0)$ which is a solution of the outer Cauchy problem 
\begin{equation}
    \begin{cases}
        z''(t)=-\omega^2 z(t)\\
        z(0)=p_0, \ z'(0)=v_0.
    \end{cases}  
\end{equation}
Since $\partial D$ is bounded and $z_E$ is an elliptic arc, there exists a \textit{first return time} for the outer dynamics; in other words, there exists $T_E>0$ such that $z_E(T_E; p_0, v_0)\in\partial D$ and $z_E(t; p_0, v_0)\notin \overline{D}$ for any $t\in(0,T_E)$. We can then consider $(\tilde \xi, \tilde \alpha')$ as the parameters that describe, using the same rationale as before, $\left(z_E(T_E; p_0,v_0), z'_E(T_E; p_0,v_0)\right)$. At this point, we have a trajectory that hits the boundary and must be refracted: following Eq.\eqref{eq:snell} with $z=\gamma(\tilde\xi)$ and $\alpha_E=\tilde\alpha'$, one can find $\alpha_I$ and the corresponding initial conditions for the outer arc $(\tilde p,\tilde v)$ (without giving the analytical formulae, we refer again to Figure \ref{fig:first return}), to define the inner arc $z_I(t; \tilde p,\tilde v)$ as the solution of the ODE
\begin{equation}
    \begin{cases}
        z''(t)=-\frac{\mu}{|z(t)|^3}z(t)\\
        z(0)=\tilde p, \ z'(0)=\tilde v.
    \end{cases}
\end{equation}
Note that Eq.\eqref{eq:snell} implies automatically that the initial velocity satisfies the inner energy equation, that is, $|\tilde v|^2/2-(\mathcal E+h)-\mu/|\tilde p|=0$. Again, given that $z_I$ is an unbounded Keplerian hyperbola, there exists a first return instant $T_I$ on $\partial D$: we can then take the outer arc's final conditions $(z_I(T_I; \tilde p,	\tilde  v), z'_I(T_I; \tilde p, \tilde v))$ and refract the inner velocity to obtain conditions $(p_1, v_1)$ such that it holds $p_1\in\partial D$, $v_1$ points outwards the domain $D$ and $|v_1|^2/2-\mathcal E+\omega^2|p_1|^2/2=0$. The final conditions $(p_1, v_1)$can be again parametrised through a pair of one dimensional quantities $(\xi_1, \alpha_1)$, and can be used as initial conditions for a new outer arc: the above machinery can be then iterated to obtain a new concatenation.\\
The map 
\begin{equation}
    \mathcal F: I\times (-\pi/2, \pi/2)\to I\times (-\pi/2, \pi/2), \quad (\xi_0, \alpha_0)\mapsto(\xi_1,\alpha_1) 
\end{equation}
is called \textit{first return map}, and can be used to describe the dynamics of our billiard in the phase space, parametrised by the variables $(\xi,\alpha)$, every time a complete concatenation of outer and inner arc is performed. \\
At the moment, we don't make any assumption $D$, except for the smoothness of its boundary; on the other hand, as ordinary in billiards theory, the dynamical properties of the system (good definition of $\mathcal F$, existence of equilibrium/periodic orbits, stability of the latters, integrable rather than chaotic behaviour) depend crucially on the geometric features of $\partial D$. Sections \ref{ssec:bil primo_art}, \ref{ssec:bil secondo_art} and \ref{ssec:bil terzo_art} aim to describe, from different points of view, such complex interdependence between the geometry of $D$ and the dynamics of our billiard.

\subsection{Equilibrium trajectories, stability and bifurcations}\label{ssec:bil primo_art}

\begin{figure}
    \centering
    \includegraphics[width=0.3\linewidth]{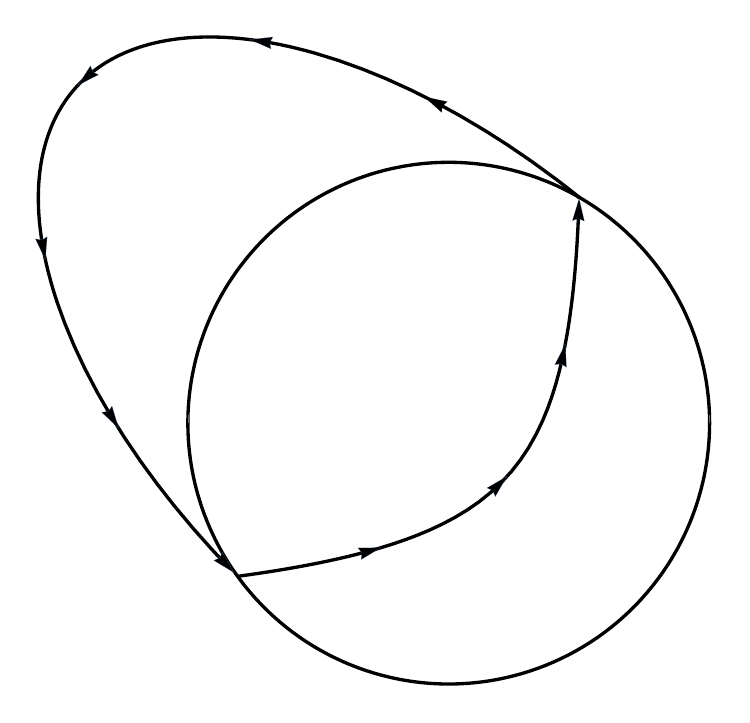}
    \quad
    \includegraphics[width=0.3\linewidth]{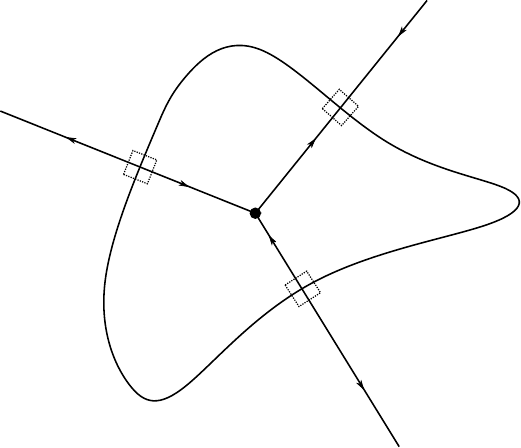}
    \caption{Equilibrium trajectories in the refraction billiard. Left: concatenation of non-homothetic inner and outer arc that refract one in the other; the existence of this kind of trajectory will be proved analytically, in the case of a circular domain, in Proposition \ref{prop:cerchio rot}. Right: examples of homothetic equilibrium trajectories. Figures taken from \cite{deblasiterraciniOnSome, barutello2023chaotic}. }
    \label{fig:nonhom equilibrium}
\end{figure}

\begin{figure}
    \centering
    \includegraphics[width=0.3\linewidth]{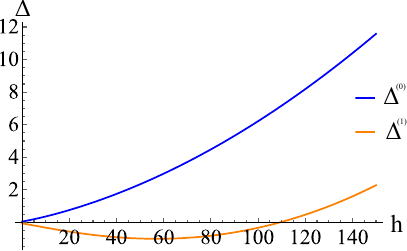}
    \quad\qquad \includegraphics[width=0.4\linewidth]{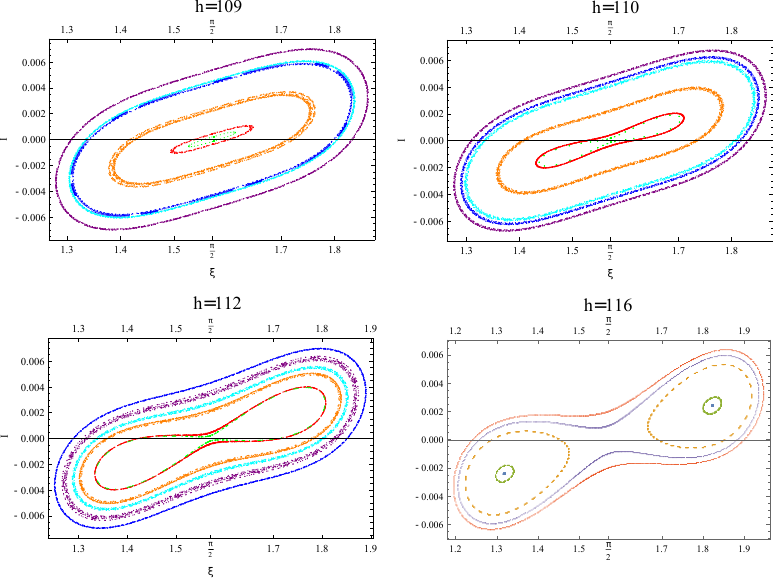}
    \caption{Left: values of the discriminants related to the central configuration $\bar\xi^{(0)}$ and $\bar\xi^{(1)}$ for $\mathcal E=2.5$, $\omega=\sqrt{2}$, $\mu=2$, $e=0.1$ and increasing values of $h$. Right: orbits of the first return map $\mathcal F$ in a neighborhood of $\bar\xi^{(1)}$ for the same parameters' value. Figures taken from \cite{deblasiterracinirefraction}.  }
    \label{fig:delta}
\end{figure}

Whenever a new dynamical system is taken into account, it is quite natural to start its analysis by searching for its equilibrium trajectories, as well as investigate their stability, using the tools of nonlinear analysis. In the formalism of the first return map, equilibrium trajectories of the two dimensional system correspond to \textit{fixed points} for $\mathcal F$ (see for example Figure \ref{fig:nonhom equilibrium}). 
In the case of the refraction billiard, there is a particular class of equilibrium trajectories, called \textit{homothetic}, whose existence is ensured provided very simple conditions on the boundary are verified: such trajectories result to be of paramount importance for the analysis of the model in many different circumstances. \\
Let us suppose to have $\bar \xi\in I$ such that: 
\begin{equation}\label{eq:cc cond}
\begin{aligned}
    &(1)\text{ the position vector for the origin to }\gamma(\bar\xi)\text{ is orthogonal to }\partial D\text{, namely, }\gamma(\bar\xi)\bot\dot\gamma(\bar\xi); \\ 
    &(2)\text{ the segment from the origin to }\gamma(\bar \xi)  \text{ intersects }\partial D   \text{ only once.}
\end{aligned}
\end{equation}
If this happens, it is easy to show that the straight half-line from the origin in the direction of $\gamma(\bar\xi)$ is invariant under both inner and outer dynamics (note that, in the case of the inner dynamics, a Levi-Civita approach to regularise the collision at the origin has been employed, cfr. \cite{Levi-Civita, deblasiterracinirefraction}), and it is not deflected by Snell's law (see Figure \ref{fig:nonhom equilibrium}, right). Along the direction defined by $\gamma(\bar\xi)$ it is then possible to construct an equilibrium trajectory, called \textit{homothetic}, which corresponds to the \textit{homothetic} fixed point $(\bar\xi,0)$ of the first return map $\mathcal F$. We highlight that, although a bouncing after the collision with the central mass might seem odd from a physical point of view, the analytic continuation of inner homothetic arcs after the collision allows to study in details the local dynamics around the singularity, giving a clear portrait of orbits which are close to the collision, and then physically relevant. It is easy to observe that condition $(1)$ in \eqref{eq:cc cond} is equivalent to require that $\bar\xi$ is a critical point for the function $|\gamma(\cdot)|$, while condition $(2)$ can be described as a \textit{star-convexity} property of the domain $D$ with respect to the direction of $\gamma(\bar\xi)$. In the following, we  will refer to parameters $\bar\xi$ as in \eqref{eq:cc cond} as \textit{central configurations.}\\
The first return map $\mathcal F$ is clearly infinitely-many well defined for any point $(\bar\xi, 0)$, with $\bar\xi$ central configuration; actually, it is possible to prove that the good definition of $\mathcal F$ holds for sure \textit{locally around }$(\bar\xi,0)$. 
\begin{proposition}\label{prop:buona definizione primo ritorno}
    Let us suppose that $\gamma$ is at least $C^1$, and let $\bar\xi$ be a central configuration. Then, there exist two positive constants $\delta, \epsilon>0$ such that the first return map $\mathcal F: (\xi_0,\alpha_0)\mapsto (\xi_1,\alpha_1)$ is well defined and differentiable in $[\bar\xi-\delta,\bar\xi+\delta]\times[-\epsilon, \epsilon]$. 
\end{proposition}
The proof of Proposition \ref{prop:buona definizione primo ritorno} relies essentially in showing the existence and uniqueness of the inner and outer arcs for any initial condition sufficiently close to $(\bar\xi,0)$ in the phase space and, in the case of the inner dynamics, the transversality of such arcs; the details are given in \cite[Sections 1.3, 1.4]{deblasiterracinirefraction}. We stress that, although well defined, at this stage we do not have an explicit expression for $\mathcal F$, even in a neighborhood of the homothetic fixed point, since such expression depends on $\gamma$. \\
Since $\mathcal F$ is locally differentiable around homothetic fixed points, it is natural to continue their analysis by investigating their linear stability, asking whether it depends on the geometrical properties of $\partial D$ around $\gamma(\bar\xi)$ as well as on the physical parameters $\mathcal E, h, \omega, \mu$. Such analysis can be carried on by considering the \textit{Jacobian matrix} of $\mathcal F$ in such points, given by
\begin{equation}
    D\mathcal F(\bar\xi, 0)=
    \begin{pmatrix}
        \frac{\partial \xi_1}{\partial\xi_0}(\bar\xi, 0) & \frac{\partial \xi_1}{\partial\alpha_0}(\bar\xi, 0) \\ 
        \frac{\partial \alpha_1}{\partial\xi_0}(\bar\xi, 0) & \frac{\partial \alpha_1}{\partial\alpha_0}(\bar\xi, 0) 
    \end{pmatrix}. 
\end{equation}
Although the explicit expression of $\mathcal F$ is not known, using the implicit function theorem and knowing the analytic expression of the homothetic solutions it is possible to obtain a closed formula 
for its Jacobian: such expression is given by 
\begin{equation}\label{eq:Jacob}
\begin{aligned}
       & D\mathcal F (\bar\xi, 0)=
        \begin{pmatrix}
            F_{11}&F_{12}\\
            F_{21}&F_{22}, 
        \end{pmatrix}\\
    &F_{11}=1+\frac{2\epsilon_E+\epsilon_I}{\tilde I}+\frac{\epsilon_E\left(\epsilon_E+\epsilon_I+\tilde I\right)}{\tilde E\tilde I}\\
    &F_{12}=\sqrt{V_E(\gamma(\bar\xi))}\left(\frac{1}{\tilde I}+\frac{1}{\tilde E}\right)+\sqrt{V_E(\gamma(\bar\xi))}\frac{\epsilon_E+\epsilon_I}{\tilde E\tilde I}\\
    &F_{21}=\frac{2\epsilon_E\left(\epsilon_I+\tilde I\right)+\epsilon_I\left(\epsilon_I+2\tilde I\right)}{I_0\sqrt{V_E(\gamma(\bar\xi))}}+\frac{\epsilon_E\left[\epsilon_E\left(\epsilon_I+\tilde I\right)+\epsilon_I(\epsilon_I+2\tilde I)\right]}{\tilde E\tilde I\sqrt{V_E(\gamma(\bar\xi))}}\\
    &F_{22}=1+\frac{\epsilon_E}{\tilde E}+\frac{\epsilon_I\left(2\tilde I+\epsilon_I+\tilde E+\epsilon_E\right)}{\tilde E\tilde I}\\
    &\tilde E=\frac{\mathcal E}{2|\gamma(\bar\xi)|\sqrt{V_E(\gamma(\bar\xi))}}, \quad \epsilon_E=\left(|\gamma(\bar\xi)|k(\bar\xi)-1\right)\frac{\sqrt{V_E(\gamma(\bar\xi))}}{|\gamma(\bar\xi)|}\\
    &\tilde I=-\frac{\mu}{4|\gamma(\bar\xi)|^2\sqrt{V_I(\gamma(\bar\xi))}}, \quad \epsilon_I=-\left(|\gamma(\bar\xi)|k(\bar\xi)-1\right)\frac{\sqrt{V_I(\gamma(\bar\xi))}}{|\gamma(\bar\xi)|}
    \end{aligned}
\end{equation}
where $k(\bar\xi)$ denotes the curvature of $\gamma$ in $\bar\xi$ (see \cite{docarmo2016differential}). The analytic expression of $D\mathcal F(\bar\xi,0)$ is quite complicated, but it is easy to notice as it depends on both the geometric properties of $\gamma$ up to second order and the physical parameters $\mathcal E, \omega, h, \mu$. Let us note that, when $D$ is a circle centered at the origin with radius $R=|\gamma(\bar\xi)|$, its curvature is always equal to $1/R$: in such case, the terms $\epsilon_E$ and $\epsilon_I$ disappear, and the Jacobian reduces simply to the identity matrix. This fact is not surprising, as it is consistent with the fact that the circular refraction billiard represents an \textit{integrable} and \textit{highly degenerate} case (see also Section \ref{ssec:bil secondo_art}): as a matter of fact, on a circular domain every radial initial condition (i.e. any point $(\xi,0)$ in the $(\xi,\alpha)-$plane) defines a homothetic equilibrium trajectory, so that the homothetic fixed points are not isolated anymore, and form instead a straight line of fixed points in correspondence of $\alpha=0$. \\
As for the general case, one can notice that the curvature of $\gamma$ plays a role only in the $\epsilon_{E/I}$ terms: for this reason, such terms can be considered \textit{corrections} induced by the geometry of $\gamma$ with respect to the circular case. \\
The linear stability of $(\bar\xi,0)$ as fixed point of $\mathcal F$ can be inferred by the eigenvalues of $D\mathcal F(\bar\xi,0)$ (see \cite{devaneyhirschsmale}), which we call $\lambda_1$ and $\lambda_2$. Since $\mathcal F$ is area preserving, it holds that $\det\left(D\mathcal F(\bar\xi,0)\right)=\lambda_1\lambda_2=1$, and, whenever $\lambda_1=\lambda_2^{-1}\in\R$, the fixed point is an unstable saddle; on the contrary, if the two eigenvalues are complex and conjugated, the homothetic fixed point is a stable center. To distinguish between the two cases, in the two-dimensional case a simple criterion can be adopted: denoted by $\Delta$ the discriminant of the characteristic polynomial associated to $D\mathcal F(\bar\xi,0)$, one has that 
\begin{itemize}
    \item if $\Delta>0$, then $(\bar\xi,0)$ is a saddle; 
    \item if $\Delta<0$, then $(\bar\xi,0)$ is a center. 
\end{itemize}
The case $\Delta=0$, corresponding to $\lambda_1=\lambda_2=1$, is highly degenerate: this is what happens for example in the circular case, and, in general, nothing can be said on the linear stability. \\
Starting from Eq.\eqref{eq:Jacob}, it is possible to give an explicit formula for the discriminant $\Delta$, which is given by 
\begin{equation}\label{eq:delta}
    \begin{aligned}
        &\Delta=ABCD\\
        &A=\frac{16}{\mathcal E\mu}\left(\sqrt{V_I(\gamma(\bar\xi))}-\sqrt{V_E(\gamma(\bar\xi))}\right)\left(|\gamma(\bar\xi)|k(\bar\xi)-1\right)\\
        &B=\mathcal E-\left(|\gamma(\bar\xi)|k(\bar\xi)-1\right)\left(\sqrt{V_E(\gamma(\bar\xi))}-\sqrt{V_E(\gamma(\bar\xi))}\right)\sqrt{V_E(\gamma\xi)}\\
        & C=-\mu\sqrt{V_E(\gamma(\bar\xi))}+2|\gamma(\bar\xi)|B\sqrt{V_I(\gamma(\bar\xi))}\\
        &D=\mu+2|\gamma(\bar\xi)|\left(|\gamma(\bar\xi)|k(\bar\xi)-1\right)\sqrt{{V_I(\gamma(\bar\xi))}}\left(\sqrt{V_I(\gamma(\bar\xi))}-\sqrt{V_E(\gamma(\bar\xi))}\right). 
    \end{aligned}
\end{equation}
The sign of $\Delta$ can be investigated numerically whenever one has an explicit expression for the curve $\gamma$: in the following, we propose a thorough illustration of the elliptic case, which, in the framework of the mathematical billiards, represents a case study of great importance (see for example \cite{takeuchi2021conformal, kaloshinsorrentinolocalbirkhoff}). \\
\paragraph{\textbf{The elliptic case: analytical and numerical results}}
To give a practical example on how Eq.\eqref{eq:delta} can be used to give exact information on the stability of the homothetic equilibrium trajectories (and, in some cases, on the overall properties of the first return map), let us suppose that $\gamma$ describes an ellipse with center in the origin, semimajor axis equal to $1$ and eccentricity $e$, that is,
\begin{equation}\label{eq:gamma}
    \gamma(\xi)=(\cos{\xi}, b\sin{\xi}), \ \xi\in[0,2\pi], \ b=\sqrt{1-e^2}. 
\end{equation}
In this case the only four homothetic trajectories are in correspondence of $\bar\xi^{(0)}=0,\ \bar\xi^{(1)}=\pi/2,\ \bar\xi^{(2)}= \pi$ and $\bar\xi^{(3)}=3\pi/2$, and they are pairwise symmetric. For any of the corresponding homothetic points, it is then possible to compute $D \mathcal F(\bar\xi^{(i)}, 0)$, $i=0,\dots,3$, and, consequently, the discriminants $\Delta^{(0)}, \dots,\Delta^{(3)}$. The explicit expressions of these quantities, as well as rigorous asymptotic analysis, is provided in \cite[Section 1.6]{deblasiterracinirefraction}; here, we limit ourselves to an example, which is of particular significance to show the consistency between the analytical tools and the numerical results. Let us take for example the numerically computed values of $\Delta^{(0)}$ and $\Delta^{(1)}$ displayed in Figure \ref{fig:delta}, left, where we fixed $\mathcal E=2.5, \ \omega=\sqrt{2},\ \mu=2,\ e=0.1$, and the inner energy $h$ varies in $[0,150]$. 
It is clear that, while the homothetic in $\bar\xi^{(0)}=0$ is always a saddle, the stability of $\bar\xi^{(1)}=\pi/2$ changes when $h$ increases: in the literature, as for example in \cite{devaneyhirschsmale}, phenomena where the dynamical properties of a map (stability of the fixed points, number of the latters, etc.) are referred to as \textit{bifurcations}.\\
We can compare Figure \ref{fig:delta}, left, with Figure \ref{fig:delta}, right, which shows the Poincar\'e map (that is, the representation in the $(\xi,\alpha)-$plane of the iterates for $\mathcal F$ for different initial points) in a neighborhood of $(\pi/2,0)$ for different values of $h$ close to the value of $h$ at which $\Delta^{(1)}$ changes sign. One can clearly see as the fixed point, which initially is a center, changes its stability, becoming a saddle, and, for increasing values of $h$, leading to the formation of a new non-homothetic, $2-$periodic point. 

\begin{figure}
    \centering
    \includegraphics[width=0.4\linewidth]{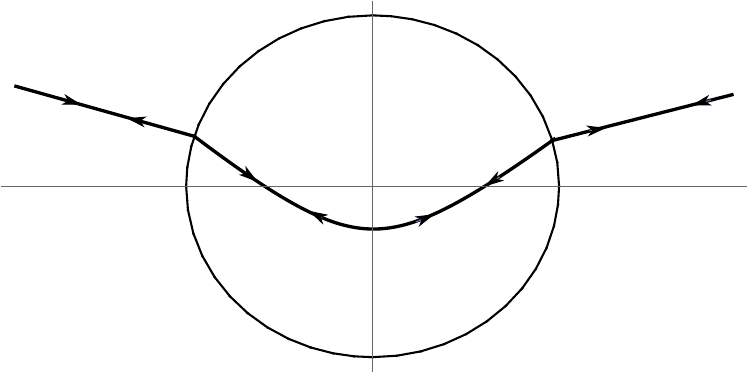}
    \qquad
    \begin{overpic}[width=0.4\linewidth]{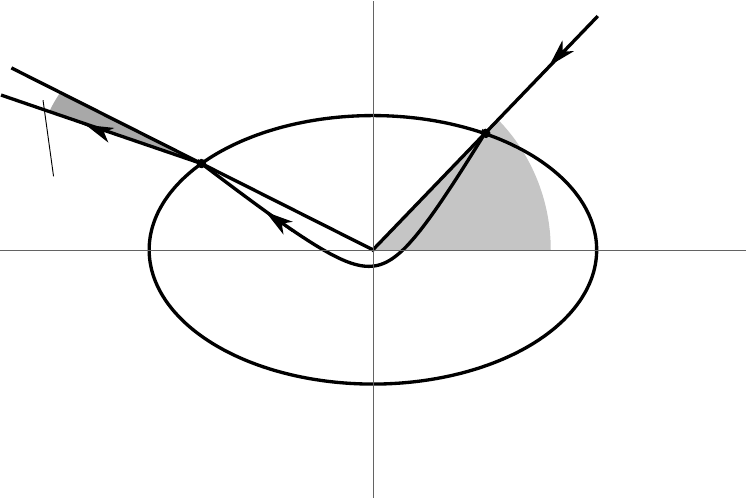}
    	\put(65,38){\tiny$\theta$}
    	\put(7,39){\tiny$\delta$}
    	\end{overpic}
    \caption{Left: example of brake two-periodic trajectory. Right: construction of the free fall map: given a direction defined by $\theta$, it returns the angle $\delta$ between the refracted outer arc and the corresponding radial direction. Figures taken from \cite{deblasiterracinirefraction}. }
    \label{fig:brake}
\end{figure}

The homothetic equilibrium trajectories analysed up to now are of great importance also for the further analysis, and in particular in Section \ref{ssec:bil terzo_art}; nevertheless, there exists another class of (two-periodic) equilibrium trajectories whose existence can be derived by purely analytic arguments. This is the case, for example, of the two periodic \textit{brake} orbits composed by a pair of outer homothetic arcs connected by an inner hyperbola (see Figure \ref{fig:brake}, left). 
Such kind of trajectories can appear whenever an inner arc refracts in both sides in radial directions, and their existence can be showed analytically by means of a \textit{free fall method}. The general idea under the free fall in this case is to construct a univariate function that, given the initial conditions corresponding to an homothetic outer arc, follows the generated trajectory until it exits again from $D$ and returns the angle between the subsequent outer arc and the radial direction in the exit point (Figure \ref{fig:brake}, right). In this way, provided the above function (called  \textit{free fall map}) is well defined (and, possibly, differentiable), the search for two periodic brake orbits translates in searching for  zeroes of a continuous function. \\
The good definition of the free fall map follows from a more general geometric property of elliptic boundaries. Since this result is interesting also by itself, we write it down in the following Proposition. 
\begin{proposition}{\cite[Proposition 6.3]{deblasiterracinirefraction}}
    Let $D$ be an elliptic domain whose boundary is parametrised as in \eqref{eq:gamma}, with $e\in[0,1/\sqrt{2})$. Then, for any $\mathcal E, h, \mu>0$, every Keplerian arc of energy $\mathcal E+h$ and mass parameter $\mu$ intersects $\partial D$ at most in two points. 
\end{proposition}
Let us remark that $1/\sqrt{2}\simeq 0.707$: the above Proposition holds then for a wide class of ellipses, not necessarily close to a circle. Whenever $e\in[0,1/\sqrt{2})$ the free fall map can be proved to be well defined,  the existence of brake two-periodic trajectories for suitable values of the physical parameters can be proved. 
\begin{theorem}{\cite[Theorem 6.4]{deblasiterracinirefraction}}
    Fixed every $\mathcal E,\omega>0$ such that $\omega^2>\mathcal E$ and any ellipse with the center at the origin, semimajor axis equal to $1$ and $e\in[0,1/\sqrt{2})$, if $\mu$ and $h$ are sufficiently large, then the first return map admits at least four two-periodic brake trajectories.
\end{theorem}
The results proposed until now hold in a local sense for    quite generic domains and, in a more global setting, in the special case of elliptic domains. In Section \ref{ssec:bil secondo_art} we will try to provide \textit{global} results for a more general class of domains: it is the case of the \textit{close to circle} ones, which will be analysed through the powerful tools brought forth by perturbation theory.

\subsection{Quasi-circular domains: a perturbative approach}\label{ssec:bil secondo_art}
It is already clear from the discussion in Section \ref{ssec:bil primo_art} that the shape of the domain $D$ is of fundamental importance to infer the properties of the billiard map; in the circular case, this become evident, as the central symmetry of the domain has radical consequences on $\mathcal F$, which will be discussed in the next paragraph. The system, in such case, results to be \textit{globally well defined, completely integrable} and admits orbits of \textit{any rotation number} within a certain interval.   \\
When the domain $D$ is sufficiently close to a circle, one can ask whether some of these properties (as for example the good definition or the existence of some particular orbits) are still maintained: in this Section we aim to answer this question, taking advantage of  tools coming from perturbation theory and general facts holding for area preserving maps (for a wide dissertation on the subject, see \cite{gole2001symplectic}). Such instruments require a definition of the analytical framework we are working in which is a bit deeper than the one described in the previous Sections, in particular involving the so-called \textit{generating function} (see \cite{gole2001symplectic}), that now we will define in our specific case. \\
Let us assume again that the boundary of our domain can be parametrised by a curve $\gamma$: for the sake of simplicity, we will assume that $\gamma$ is $2\pi$-periodic, and, with an abuse of notation, we still denote with $\gamma$ the periodic extension of the curve, namely, $\gamma:\R_{/2\pi Z}\to \R^2$, where $\R_{/2\pi\Z}$ denotes the $2\pi-$periodic torus. Let us now take the function
\begin{equation}	\label{eq:generatrice}
	\begin{aligned}
	&G:\R_{/2\pi\Z}\times\R_{/2\pi\Z}	\to \R,\\
	&G(\xi_0,\xi_1)=G_E(\xi_0,\tilde\xi)+G_I(\tilde\xi, \xi_1)=\mathcal{L}_E(\gamma(\xi_0),\gamma(\tilde\xi))+\mathcal L_I(\gamma(\tilde\xi),\gamma(\xi_1)), 
	\end{aligned}
\end{equation}
where $\mathcal L_E$ and $\mathcal L_I$ are the outer and inner Jacobi distance as  defined in Eq.\eqref{eq:Jacobi dist}. By means of the implicit function theorem,  the parameter $\tilde \xi$ can be expressed as a function of $\xi_0$ and $\xi_1$ from the relation
\begin{equation}
	\frac{\partial}{\partial \xi}\left(\mathcal{L}_E(\gamma(\xi_0), \gamma({\xi}))+\mathcal{L}_I(\gamma({\xi}), \gamma(\xi_1))\right)_{\xi=\tilde\xi}=0;   
\end{equation}
provided the following non degeneracy condition holds
\begin{equation}\label{eq:ndg1}
	\partial_{\xi}\left(G_E(\xi_0,\xi)+G_I(\xi,\xi_1)\right)\neq0. 
\end{equation}
Recalling the variational interpretation of Snell's law, one can notice that, given two points $p_0=\gamma(\xi_0)$ and $p_1=\gamma(\xi_1)$, the generating function $G$ returns the Jacobi length of the concatenation that connects $p_0$ to $p_1$ with an outer and inner arc, and trespasses the boundary precisely at the point $\tilde p=\gamma(\tilde\xi)$ which ensures that the refraction law is satisfied by the arcs. The good definition of $G$, ad well as its differentiability, is not always guaranteed, and depends on $D$; more precisely, it is associated to the existence and uniqueness of outer and inner arcs connecting any pair of points on $\partial D$, and must be verified case by case. \\
Generating functions are commonly used when dealing with billiards (see also \cite{Tabbook}), since the first return map, in a suitable set of canonical \textit{action-angle} variables, can be implicitly expressed in terms of derivatives of $G$. In particular, one can define the \textit{canonical actions } conjugated with the parameters $\xi_0,\xi_1$: 
\begin{equation}\label{eq:azioni}
	I_0(\xi_0,\xi_1)=-\partial_{\xi_0}G(\xi_0, \xi_1), \quad I_1(\xi_0,\xi_1)=\partial_{\xi_1}G(\xi_0,\xi_1); 
\end{equation}
such quantities, which in the following will replace the angles $\alpha_0, \alpha_1$ in the construction of a first return map, have in turn a geometrical interpretation, given by (see \cite{deblasiterraciniOnSome})
\begin{equation}\label{eq:azioni alpha}
	I_0=\sqrt{V_E(\xi_0)}\sin\alpha_0, \quad I_1=\sqrt{V_I(\gamma(\xi_1))}\sin\alpha_1. 
\end{equation}
Eq.\eqref{eq:azioni} translates in the fact that, whenever the initial and final point of a concatenation are known, the initial and final actions $I_0$ and $I_1$ (and, as a consequence, the angles $\alpha_0$ and $\alpha_1$) can be computed from the generating function's derivatives. Starting from this, it is possible to reconstruct the first return map in terms of the variables $(\xi_0,I_0)$ by means again  of the implicit function theorem: whenever 
\begin{equation}\label{eq:ndg2}
	\partial_{\xi_1}\left(I_0 + \partial_{\xi_0}G(\xi_0,\xi_1) \right)\neq 0,
\end{equation}
one can invert the first relation in \eqref{eq:azioni} to obtain $\xi_1$ as a function of $\xi_0$, and $I_0$, and then define the first return map\footnote{For the sake of consistency with Section \ref{ssec:bil modello} and with a slight abuse of notation, the first return map in terms of action-angle variables will be still called $\mathcal F$. } as 
\begin{equation}
	\mathcal F: (\xi_0, \alpha_0)\mapsto (\xi_1(\xi_0,I_0), I_1(\xi_0,I_0))= (\xi_1(\xi_0,I_0), I_1(\xi_0,\xi_1(\xi_0,I_0))). 
\end{equation}
The good definition of the first return map in suitable regions of $\R_{/2\pi\Z}\times \R$ is ensured whenever $G_E$ and $G_I$ are well defined and the nondegeneracy conditions \eqref{eq:ndg1} and \eqref{eq:ndg2} hold: such hypotheses will be verified case by case, possibly using different techniques. \\

\begin{figure}
	\centering
	\includegraphics[height=.17\textheight]{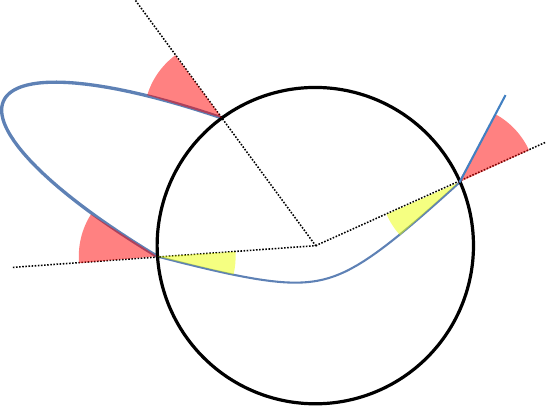}\qquad
	\begin{overpic}[height=.17\textheight]{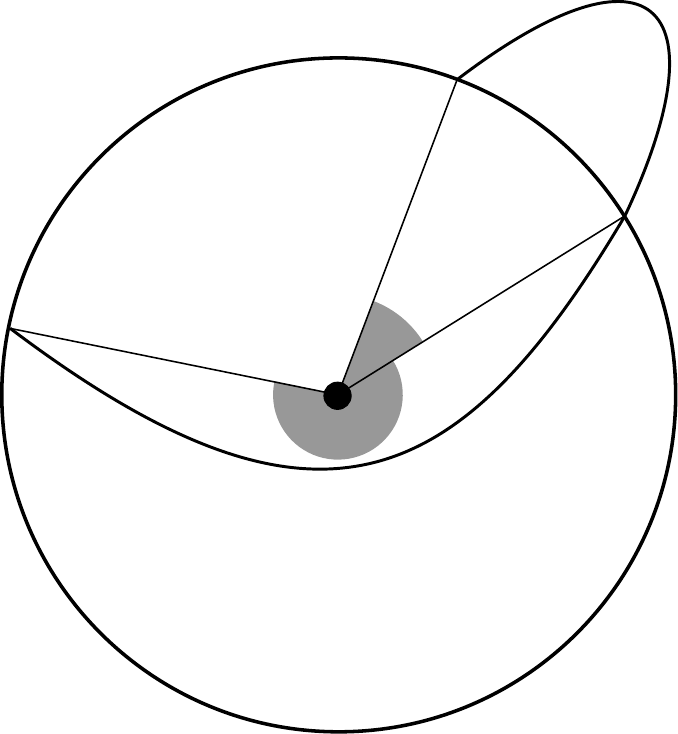}
		\put(60,63){\tiny$\theta_E$}
		\put(45,40){\tiny$\theta_E$}
	\end{overpic}
	\caption{Left: conservation of the angle $\alpha$ in the circular case. Right: Outer and inner shift in the circular case. By virtue of the central symmetry of the overall system, they depend only on $I_0$ and not on $\xi_0$. }
	\label{fig:cerchio}
\end{figure}

\begin{figure}
	\centering
	\includegraphics[width=.4\linewidth]{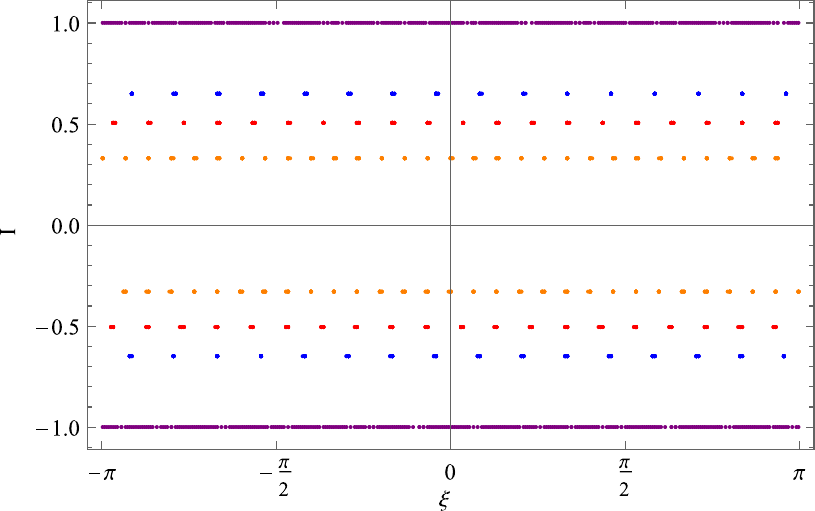}
	\caption{Orbits of the refraction galactic billiard in the circular case, taking into account the physical parameters' values $\mathcal E=7, \omega^2=3, h=2$ and $\mu=15$. The orbit lie on horizontal invariant lines with $I=const$, and could be either periodic (the dotted lines) or cover densely the line (examples are the continuous lines on the top and bottom of the figure). A particular case is given by the invariant line $I=0$, which is covered in homothetic fixed points (see Section \ref{ssec:bil primo_art}). Figure taken from \cite{deblasiterraciniOnSome}. }
	\label{fig:rot num}
\end{figure}

\paragraph{\textbf{The circular case}}The final aim of the investigation presented in the current Section is to provide dynamical results holding for the billiard map induced over a domain which is quasi circular. To do it, we will adopt a perturbative point of view, taking as a unperturbed case the circle and then applying slight modifications to the boundary. To do this, a careful analysis on the map whenever the domain is circular is in order. \\
Let us suppose that $D$ is a circle of radius $1$, whose boundary is parametrised by $\gamma(\xi)=(\cos\xi,\sin\xi)$, $\xi\in\R_{/2\pi\Z}$: in this case, both the potentials and the boundary are centrally symmetric, and, geometrically, this translates in the conservation of the angle $\alpha$ after every concatenation of outer and inner arc (see Figure \ref{fig:cerchio}, left). Taking advantage of this fact it is possible to give an explicit formulation for the first return map in this case, given by 
\begin{equation}\label{eq:cerchio F}
	\begin{aligned}
	&\mathcal F^{(c)}: \R_{/2\pi\Z}\times (-I_c, I_c)\to\R_{/2\pi\Z}\times (-I_c, I_c), \\
	&\mathcal F^{(c)}(\xi_0, I_0)=(\xi_1(\xi_0, I_0), I_1(\xi_0,I_0))=(\xi_0+\theta_E(I_0)+\theta_I(I_0), I_0), \\
	&\theta_E(I_0)=
	\begin{cases}
		\text{arccot}\left(\frac{\mathcal E-2I_0^2}{I_0\sqrt{4\mathcal E-2(2I_0^2+\omega^2)}}\right) \quad &\text{if }I_0>0\\
		0 &\text{if }I_0=0\\
		\text{arccot}\left(\frac{\mathcal E-2I_0^2}{I_0\sqrt{4\mathcal E-2(2I_0^2+\omega^2)}}\right)-\pi \quad &\text{if }I_0<0
	\end{cases} \\
	&\theta_I(I_0)=
	\begin{cases}
		2\arccos\left(\frac{2I_0^2-\mu}{\sqrt{4(\mathcal E+h)I_0^2+\mu^2}}\right)-2\pi \quad &\text{if }I_0>0\\
		0 &\text{if }I_0=0\\
		-2\arccos\left(\frac{2I_0^2-\mu}{\sqrt{4(\mathcal E+h)I_0^2+\mu^2}}\right)+2\pi \quad &\text{if }I_0>0
	\end{cases}
	\end{aligned}
\end{equation}
where $I_c=\sqrt{\mathcal E-\omega^2/2}$. Equation \eqref{eq:cerchio F} has been obtained by means of analytical and geometrical reasonings, coming for classical results of Celestial Mechanics as well; the detailed computations can be found in \cite{deblasiterraciniOnSome}. We can observe that $\mathcal F^{(c)}$ is in the form of a \textit{shift map}, that is, a map that at every iterate operates a change in the angle which depends only on $I_0$ by keeping constant the action. This is a direct consequence of the rotational invariance of the problem in the circular case, and shows as the action, once fixed by the initial conditions, is an \textit{integral of the motion} of the system.  For this reason, we can say that, in the circular case, the first return map is \textit{completely integrable}, since it is a two-dimensional discrete map and has two conserved and independent quantities. The shift in the angle is a $C^1$ function of $I_0$, and can be splitted into two terms: $\theta_E$ represents th angle displacement after the outer arc, while $\theta_I$ is the shift in the final angle after the Keplerian inner arc (see Figure \ref{fig:cerchio}). 
We can clearly say that $\mathcal F^{(c)}$ is globally well defined in $\R_{/2\pi\Z}\times (-I_c, I_c)$; moreover, one can check that, for any $I\in(-I_c, I_c)$, one has that $\theta'_E(I)>0$ and $\theta_I'(I)<0$. \\
The first problem we want to address in the circular case is the possible presence of \textit{periodic orbits}; to do this, let us introduce the concept of rotation number (see \cite{gole2001symplectic}). 
\begin{definition}
	Given an initial point $(\xi_0, I_0)\in\R_{/2\pi\Z}\times (-I_c, I_c)$, let us denote the (backward and forward) \emph{orbit} generated by it through  $\mathcal F^{(c)}$ as the sequence  $\left((\xi_k, I_k)\right)_{k\in\Z}=\left( \left(\mathcal F^{(c)}\right)^k(\xi_0,I_0)\right)_{k\in\Z}$. Using this notation, the (forward) \emph{rotation number} associated to the orbit is given by
	\begin{equation}\label{eq:rot}
		\rho(\xi_0,I_0)=\lim_{k\to\infty}\frac{\xi_k-\xi_0}{k}. 
	\end{equation}
\end{definition}
Although the above definition holds in general, it is clear that, in the circular case, the rotation number depends only on $I_0$ and is simply given by the shift $\theta_E(I_0)+\theta_I(I_0)$. \\
Orbits with $2\pi-$rational rotation number can be periodic, in the sense that, supposing $\rho(\xi_0,I_0)=2\pi p/q$, one has a $(p,q)-$\textit{periodic orbit} such that 
\begin{equation}
	(\xi_q, I_q)=(\xi_0+2\pi p, I_0)=(\xi_0, I_0);   
\end{equation}
on the other hand, whenever $\rho(\xi_0, I_0)\notin \Q$, the corresponding orbit covers densely the invariant line $\{(\xi, I_0)\ |\ \xi\in\R_{/2\pi\Z}\}$ in the $(\xi,I)-$plane (see Figure \ref{fig:rot num}). The line $\{(\xi, 0)\ |\ \xi\in\R_{/2\pi\Z}\}$, although still invariant, represents an exception to the dichotomy between periodic (but \textit{moving}) and dense orbits: it corresponds to all the initial conditions for the homothetic fixed points (see Section \ref{ssec:bil primo_art}), and it is then covered in a continuum of points with rotation number equal to 0.
The following Proposition resumes the results obtained in the circular case in terms of existence of (periodic and non-periodic) orbits with fixed rotation number; for the analytical expression of all the threshold values, as well as the proof, one can check \cite{deblasiterraciniOnSome}. 
\begin{proposition}{\cite[Propositions 4.9 \& 4.10]{deblasiterraciniOnSome}}\label{prop:cerchio rot}
	There exists a constant $C\in(0,\pi)$, which depends on the physical parameters $\mathcal E, h, \mu$ and $\omega$, such that for every $\rho\in (-C,C)$ there exists an action value $I\in(0,I_c)$ for which, for every $\xi_0\in\R_{/2\pi\Z},$ one has $\rho(\xi_0, \pm I)=\rho$. In particular, for any $p,q\in\Z$ such that $2\pi p/q\in(-C,C)$, there exist at least two $(p,q)-$periodic orbits starting at any $\xi_0\in\R_{/2\pi\Z}$. \\
	Furthermore, for an open set of values of the physical parameters (that is, an open set in the $(\mathcal E, \omega, \mu, h)-$space opportunely defined), there exist at least two non-homothetic fixed points of $\mathcal F^{(c)}$. 
\end{proposition}
Note that the second part of Proposition \ref{prop:cerchio rot} proves analytically the existence of non-homothetic fixed points in the circular case, as the one displayed in Figure \ref{fig:nonhom equilibrium}, left. 
\paragraph{\textbf{Quasi-circular domains}} Once the analysis of the unperturbed case has been carried on, one can ask whether and under which conditions Proposition \ref{prop:cerchio rot}  continues to hold when we are dealing with domains which are close to be circular. The class of the quasi-circular domains we are taking into account is obtained by performing a \textit{radial deformation} of the boundary depending on a generic smooth function and on a parameter $\epsilon$, namely, 
\begin{equation}\label{eq:gammapert}
	\tilde \gamma:\R_{/2\pi\Z}\times [-C_{\epsilon}, C_\epsilon]\to \R^2, \quad \tilde \gamma(\xi; \epsilon)=(1+\epsilon f(\xi; \epsilon))(\cos\xi, \sin\xi), 
\end{equation}
where $f(\xi, \epsilon)$ is a one-valued function smooth in both variables and $C_\epsilon>0$ is arbitrarily large. In this way, for any $\epsilon>-1$ the parameter $\xi$ still represents the angle between the corresponding point $\tilde \gamma(\xi;\epsilon)$ and the $x-$axis.  We will refer as $D_\epsilon$ to the domain whose boundary is parametrised by $\tilde \gamma(\cdot; \epsilon)$. \\
In general, when $\epsilon\neq0$ the central symmetry which characterise the circular case breaks, and we are no longer able to provide an explicit expression for the first return map on	$\partial D$. Nonetheless, it is possible to prove an analogous of Proposition \ref{prop:cerchio rot}, by using a more powerful set of tools, coming from the general theory of area preserving maps. To take advantage of them, it is necessary to consider again the generating function as defined in \eqref{eq:generatrice}: 
\begin{equation}
	G(\xi_0,\xi_1; \epsilon)=\mathcal L_E(\tilde \gamma(\xi_0; \epsilon), \tilde \gamma(\tilde \xi; \epsilon))+\mathcal L_I(\tilde \gamma(\tilde \xi; \epsilon), \tilde \gamma(\xi_1; \epsilon)), 
\end{equation}
where again $\tilde\xi$ can be implicitly defined in neighborhoods of points $(\xi_0,\xi_1)$ for which condition \eqref{eq:ndg1} is satisfied. Restricting the domain in the actions and assuming that the perturbing function $f$ in \eqref{eq:gammapert} is regular enough, as well as $\epsilon$ small enough, on can prove that the first return map generated by $G(\cdot, \cdot; \epsilon)$ is well defined, and can be expressed as a (unknown, in principle) perturbation of  \eqref{eq:cerchio F}. 
\begin{proposition}{\cite[Lemma 4.4 \& Proposition 5.6]{deblasiterraciniOnSome}}\label{prop:primo ritorno pert}
	Suppose that $f\in C^k(\R_{/2\pi\Z})$, $k\geq2$, and define $\mathcal J$ as the set of all the values $I\in(-I_c, I_c)$ such that $\theta'_E(I)+\theta'_I(I)=0$. Then $\mathcal J$ is finite and for every $[a,b]\subset(-I_c, I_c)\setminus\mathcal J$ and $\epsilon$ sufficiently small the perturbed first return map on $D_\epsilon$, denoted with 
	\begin{equation}
		\mathcal F(\xi_0,I_0; \epsilon)=(\xi_1(\xi_0, I_0; \epsilon), I_1(\xi_0,I_0; \epsilon)),
	\end{equation}
	is well defined in $\R_{/2\pi \Z}\times [a,b]$ and of class $C^{k-2}$. Moreover, it is area preserving and it holds 
	\begin{equation}\label{eq:twist}
		\frac{\partial }{\partial I_0}\xi_1(\xi_0, I_0; \epsilon)\neq 0
	\end{equation}
	for every $(\xi_0,I_0)\in\R_{/2\pi \Z}\times [a,b]$. 
\end{proposition}
The need of restricting the action domain is related to the nondegeneracy conditions \eqref{eq:ndg1} and \eqref{eq:ndg2}. \\
In general, two-dimensional area preserving maps for which \eqref{eq:twist} holds are called \textit{twist maps}: such property will be important for the forthcoming analysis.
By continuity with respect to $\epsilon$, whenever Proposition \ref{prop:primo ritorno pert} holds the map $\mathcal F(\cdot, \cdot; \epsilon)$ can be written as a perturbation of $\mathcal F^{(c)}$ of the form
\begin{equation}
	\mathcal F(\xi_0, I_0; \epsilon)=(xi_0+\theta_E(I_0)+\theta_I(I_0)+A(\xi_0, I_0; \epsilon), I_0+B(\xi_0, I_0; \epsilon)), 
\end{equation}
where $A$ and $B$ are two unknown functions of class $C^{k-2}$ and tend to the constant zero-function whenever $\epsilon\to0$ in the $C^{k-2}-$norm. \\
Now that we have defined the first return map, as well as its domain, for quasi-circular billiards,  we can build up a proving algorithm, which, taking advantage of some general results for area preserving maps, will ensure the existence of orbits with fixed rotation number also in the perturbed setting. \\
To do this, we will use two powerful results of nonlinear analysis, namely, the \textit{Poincar\'e-Birkhoff theorem} and the \textit{Aubry-Mather} one. In general (for a more rigorous explanation, see \cite{gole2001symplectic, deblasiterraciniOnSome}), these theorems apply to an area-preserving twist homeomorphism on an annulus $\R_{/2\pi\Z}\times[a,b]$, which preserves the boundaries $\R_{/2\pi\Z}\times \{a\}$ and $\R_{/2\pi\Z}\times \{b\}$ with rotation numbers $\rho_a$ and $\rho_b$. As for  Poincar\'e-Birkhoff theorem, it focuses on the periodic case, claiming that for any $2\pi-$rational number $\rho\in(\rho_a,\rho_b)$ there are at least two periodic orbits whose rotation number is precisely $\rho$. Aubry-Mather theorem extends the result to \textit{any} real number in the interval $(\rho_a,\rho_b)$, claiming the existence of at least one orbit for any of them. \\
Before applying these results, it is necessary to construct an invariant set of $\mathcal F(\cdot,\cdot; \epsilon)$ where the map is surely well defined, area preserving and twist and whose boundaries are invariant curves for $\mathcal F$. To do this, we shall take advantage of the celebrated \textit{KAM theorem} (see \cite{moser1962invariant}), an extremely powerful result in perturbation theory that ensures the persistence of suitable invariant curves under small changes in a shift map as in \eqref{eq:cerchio F}. \\
The overall proving scheme towards the final result can be resumed as follows: 
\begin{itemize}
	\item we construct an invariant set $\mathcal K$ by means of KAM theorem. More precisely, we prove that for $\epsilon$ sufficiently small there exist two curves over the $\xi-$axis which are invariant under $\mathcal F(\cdot,\cdot; \epsilon)$ and have irrational rotation number. The set $\mathcal K$ is precisely the region of the $(\xi,I)-$plane bounded by these two curves (see Figure \ref{fig:invariante});
	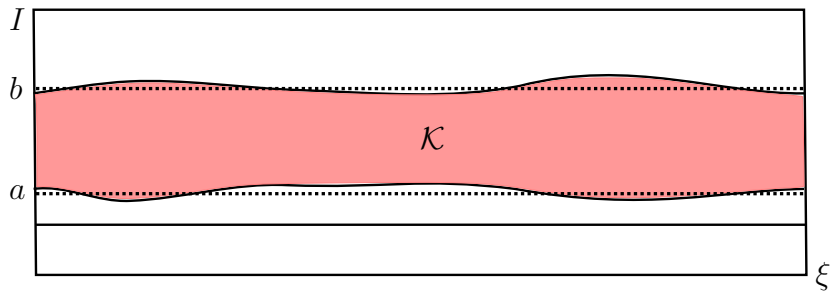
\begin{figure}
		\centering
		\begin{overpic}[width=.6\linewidth]{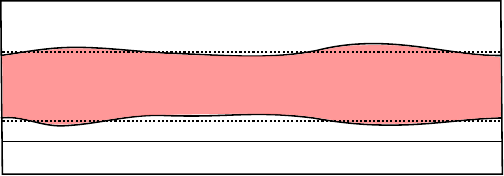}
			\put (101,-1) {\rotatebox{0}{$\xi$}}
			\put (-3,32) {\rotatebox{0}{$I$}}
			\put (-3,23) {\rotatebox{0}{$b$}}
			\put (-3,10) {\rotatebox{0}{$a$}}
			\put (50,16.5) {\rotatebox{0}{$\mathcal K$}}
		\end{overpic}
		\caption{Construction of the invariant set $\mathcal K$ for the perturbed first return map $\mathcal F(\cdot,\cdot; \epsilon)$ in the $(\xi,I)$-plane. The curved solid lines are the invariant curves for the perturbed dynamics, whose existence is ensured by KAM theorem and with rotation number $\rho_a$ and $\rho_b$. The dashed straight lines are the corresponding orbits in the unperturbed dynamics. }
		\label{fig:invariante}
	\end{figure}	
	\item with a suitable change of coordinates, we will deform $\mathcal K$ to put it in a form of an annulus $\R_{/2\pi\Z}\times[a,b]$: in the new coordinates, the perturbed map $\mathcal F$ satisfies the hypotheses of both Poincar\'e-Birkhoff and Aubry-Mather theorems; 
	\item we apply the two theorems to obtain, finally, our existence result. 
\end{itemize}

	The most delicate part in the above scheme consists in finding a regime  (namely, a region in  the $(\xi, I)-$plane) and a sufficiently small value of $\epsilon$ for KAM theorem to be applied. As a first point, we shall choose carefully the curves which we want to preserve under perturbation: they must have a \textit{Diophantine} rotation number (see \cite{moser1962invariant} and \cite{deblasiterraciniOnSome}). We also stress that, in view of Proposition \ref{prop:primo ritorno pert}, the above procedure can be applied to any connected component of $(-I_c, I_c)\setminus\mathcal J$, so the multiplicity of the found orbits can change accordingly. 
We present now the final result of this Section without any further detail on the proof: nonetheless, we invite the interested reader to go through the application of all the proposed techniques in \cite{deblasiterraciniOnSome}. 
\begin{theorem}{\cite[Theorem 5.20]{deblasiterraciniOnSome}}\label{thm:rot}
	Let us suppose that $f$, defined as in \eqref{eq:gammapert}, is of class $C^k$, with $k>5$. Let us define $E_1, \dots, E_N$ as the connected components of $(-I_c, I_c)\setminus\mathcal J$, and, fixed any $[a_i,b_i]\subset E_i$ for every $i=1, \dots, N$, set $\theta_-^i$ and $\theta_+^i$ as the rotation numbers for the unperturbed dynamics for $I_0=a_i$ and $I_0=b_i$ (for simplicity, let us assume that $\theta_-^i<\theta_+^i$). For any $i=1, \dots, N$, let us fix $\rho_\pm^i$ two Diophantine numbers such that $\theta_-^i<\rho_-^i<\rho_+^i<\theta_+^i$. Then there exists $\bar\epsilon>0$ such that, for every $\epsilon \in\R$, $|\epsilon|<\bar\epsilon$, and every $\rho\in(\rho_-^i, \rho_+^i)$ there is at least $k$ orbits for the perturbed map with rotation number $\rho$, where $k$ is the number of intervals $(\rho_-^i, \rho_+^i)$, for different $i$, in which $\rho$ is contained. Moreover, if $\rho$ is $2\pi-$rational, then the orbits are $2k$ and they are periodic. 
\end{theorem}
Although a little bit technical in its set up, the core of the theorem is that, restricting suitably the set of rotation numbers and taking domains which are sufficiently regular and close to a circle, the existence of orbits with fixed $\rho$, including periodic ones, is guaranteed. \\
The existence of a wide variety of periodic orbits for a non-circular case is a highly nontrivial results, and can be interpreted as a strong hint of the presence of a \textit{complex} dynamics, which might be \textit{chaotic}: this is precisely the topic of Section \ref{ssec:bil terzo_art}, where we will search for conditions on the domain's shape under which it is possible to prove analytically the chaoticity of the model.

\subsection{The onset of chaos in galactic refraction billiards}
\label{ssec:bil terzo_art}

\begin{figure}
 	\centering
 	\includegraphics[width=0.5\linewidth]{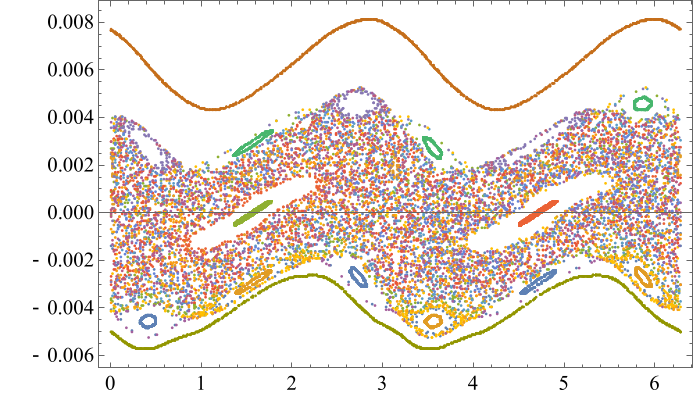}
 	\caption{Numerical evidences of chaotic phenomena in the galactic refraction billiard. The orbits correspond to the dynamics induced by an elliptic domain with eccentricity $0.05$ and the physical parameters' values $\mathcal E=20$, $\omega=1$, $\mu=0.13$ and $h=40.$ Figure taken from \cite{deblasiterracinirefraction}. }
 	\label{fig:chaos numerica}
 \end{figure}

\begin{figure}
	\centering
	\includegraphics[height=2.2cm]{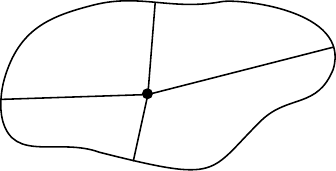}\qquad\qquad
	\includegraphics[height=2.2cm]{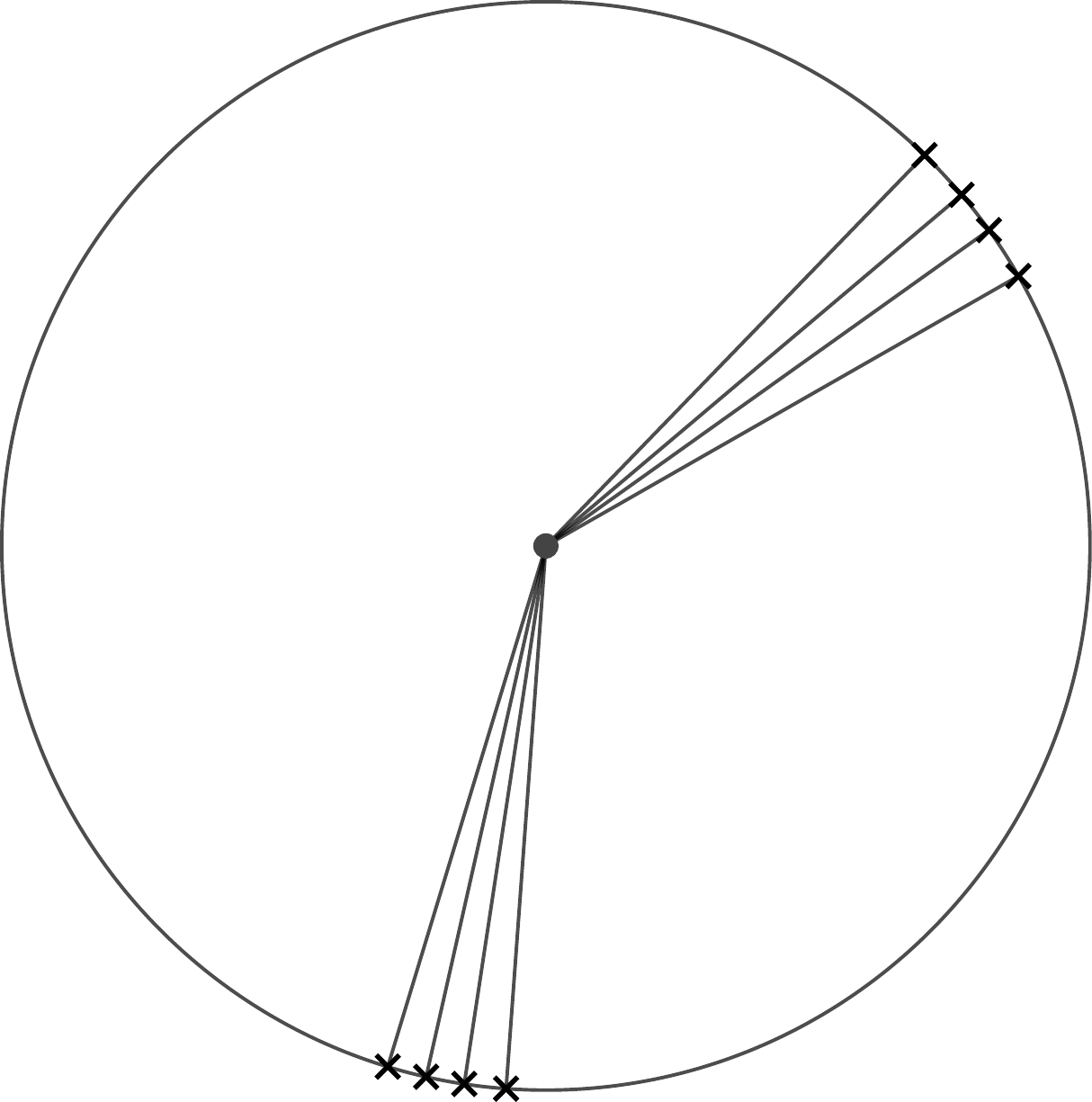}\qquad\qquad
	\includegraphics[height=2.2cm]{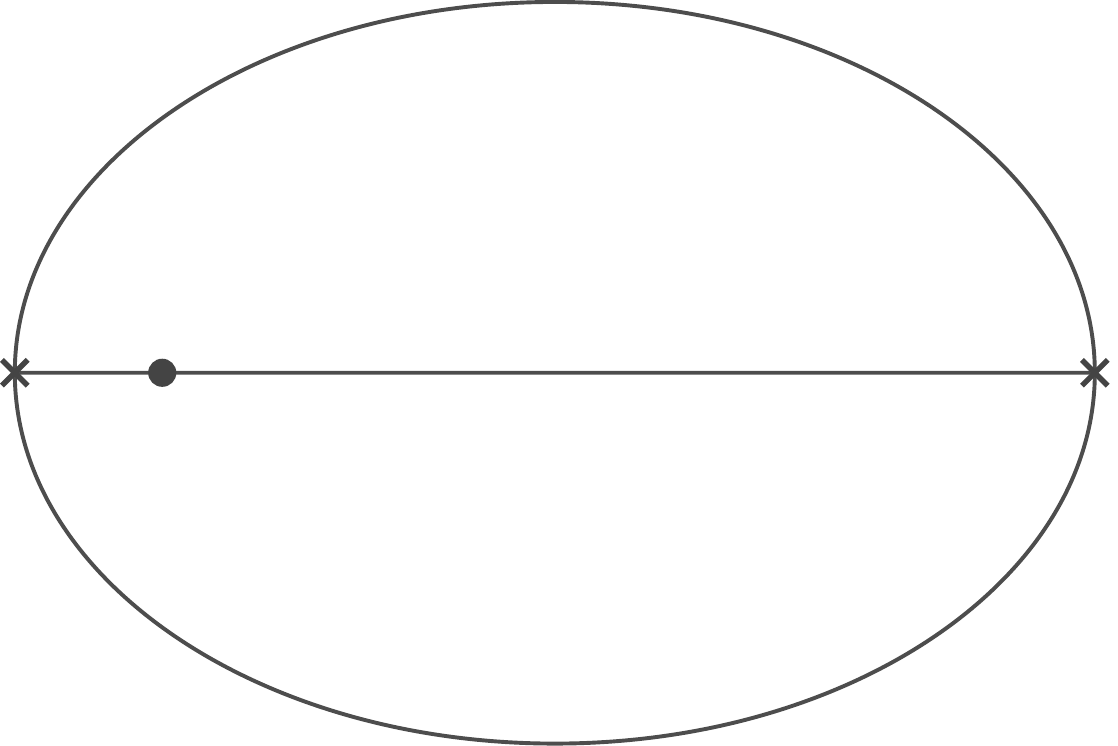}
	\caption{Right: example of admissible domain. Here, one can find two central configurations, as in Definition \ref{def:ammissibile}, which are strict maxima or minima for the distance from the origin restricted to $\partial D$ (hence nondegenerate critical points). Center and right: examples of non admissible domains. In the first case, a circle whose center is at the origin admits infinitely-many central configurations, but none of them are nondegenerate. On the other hand, an ellipse with one of the foci at the origin admits two nondegenerate central configurations, but they are antipodal. }
	\label{fig:ammissibile}
\end{figure}

 The study of the dynamics of galactic refraction billiards, carried on starting from the existence and stability of equilibrium trajectories in Section \ref{ssec:bil primo_art} and continued with the analysis of the \textit{quasi-integrable} setting induced on the model by the choice of a quasi-circular boundary, finds in this last Section its conclusion. Here, the problem addressed is the \textit{possible chaoticity} of our model, and, in particular, the existence of simple, geometrical conditions on $D$ that ensure that the system satisfies a mathematically rigorous definition of chaos. \\
 Such question, which in some sense was the first reason to motivate us in investigating the galactic refraction model, comes quite natural while observing some of the simulations provided in \cite{deblasiterracinirefraction} (see also Figure \ref{fig:chaos numerica}). At least in the particular case of an ellipse having its center at the origin, it is evident how increasing values of the inner energy $h$ lead to the appearance of diffusive orbits around the homothetic points on the $\xi-$axis, providing a clear evidence of chaotic behaviour, in a subset of the phase space. \\
 Of course, numerical simulations as the ones presented in \cite{deblasiterracinirefraction} are not enough to prove the \textit{actual} chaoticity of the model, as numerical instabilities can get in the way and lead to possibly non-accurate deductions. It is then important to go further with a \textit{rigorous} proof, to show \textit{analytically} how chaotic phenomena can occur under precise hypotheses. \\
 A first problem one has to deal with is which definition of chaos it is convenient to take into account: roughly speaking, a dynamical system is considered chaotic whenever it is sensitive to changes in the initial conditions. More precisely, when, moving from a given point, the trajectories' behaviour becomes unpredictable,  potentially covering the whole phase space. From a mathematical point of view, there are many different ways to  explicitly describe such concept (see for example \cite{hasselblattkatok}): in the current paper, we will use Devaney's definition of \textit{topological chaos}, presented in full details in \cite{Dev_book}. \\
 The main result regarding chaotic behaviour in a galactic refraction billiard, presented in \cite{deblasiterracinibarutelloChaotic}, resumes in finding a geometrical condition on the boundary, called \textit{admissibility}, which ensures the existence of a topologically chaotic subsystem for high enough inner energies. 
 \begin{definition}\label{def:ammissibile}
 	Let us take a domain $D$ containing the origin, and, taking $\gamma:I\to \R^2$ as a parametrisation of $\partial D$, suppose that it is at least $C^2$. The domain $D$ is termed 	\emph{admissible} if there are at least two nondegenerate and non-antipodal central configurations, namely, if there exist $\xi_1,\xi_2\in I$ as in Eq. \eqref{eq:cc cond} which are strict maxima or minima for the function $\|\gamma(\cdot)\|$ and such that $\gamma(\xi_1)$ and $\gamma(\xi_2)$ are not collinear with the origin (see Figure \ref{fig:ammissibile}).   
 \end{definition}
 \begin{theorem}{\cite[Theorem 4.7]{deblasiterracinibarutelloChaotic}}\label{thm:chaos}
 Let $D$ be an admissible domain as in Definition \ref{def:ammissibile}. Then, if the inner energy's parameter $h$ is high enough, the galactic refraction billiard is \emph{chaotic}, in the sense that it admits a topologically chaotic subsystem. 
 \end{theorem}
While in \cite{barutello2023chaotic} the proof of Theorem \ref{thm:chaos} is explained in details, here we limit ourselves in pointing out that such chaotic subsystem is obtained by \textit{conjugation with the Bernoulli shift}, achieved by constructing a suitable symbolic dynamics (see \cite{Dev_book}), which is well defined in a subset of the initial conditions and whenever the inner energy is high enough with respect to the outer one.\\
For the sake of completeness, let us briefly resume the main ideas behind such construction. In general, we say that a dynamical system $F:A\to A$, $A$ being the set of initial conditions, admits a symbolic dynamics whenever there exists a surjective and continuous projection map $\Pi:A\to\{1,\dots,n\}^\Z$, $n\geq1$, such that the diagram 
\[
\begin{tikzcd}
	A \arrow{r}{{F}} \arrow{d}{\Pi} & A \arrow{d}{\Pi} \\
	\{1,\dots,n\}^\Z \arrow{r}{\sigma}	& \{1,\dots,n\}^\Z
\end{tikzcd}
\]
commutes, where the function $\sigma$ is the \textit{Bernoulli shift}: given a sequence in $\{1,\dots,n\}^\Z$, $\sigma$ moves any of its elements on the right\footnote{We stress that, in \cite{Dev_book}, the definition of symbolic dynamics is given by taking into account the set of right-infinite sequences of two symbols, that is, $\{0,1\}^\N$. Without losing in generality, we took a less restrictive definition of $\sigma$ to be consistent with the construction used in the refraction billiard case.  }.\\
In practice, constructing a symbolic dynamics corresponds to \textit{encoding} the forward and backward orbits  $F$ into sequences of symbols, whose geometrical and physical meaning depends by the model itself. This fact is particularly important when $\Pi$ is bijective: in this case, we say that our system is \textit{topologically conjugated} with $\sigma$, and the one-to-one correspondence between $F$-orbits and bi-infinite sequences of symbols allows to conclude that $F$ is chaotic.   \\
In our case, our aim is to construct a topological conjugacy between the first return map $\mathcal F:(\xi_0,\alpha_0)\mapsto(\xi_1,\alpha_1)$, possibly restricted to a subsystem, and the Bernoulli shift on a suitable set of sequences. The map $\Pi$ keeps tracks of the points where subsequent concatenations of outer and inner arcs, and in particular the ones that remain close to the homothetics, intersect the boundary.   
\begin{figure}
	\centering
	\includegraphics[width=.4\textwidth]{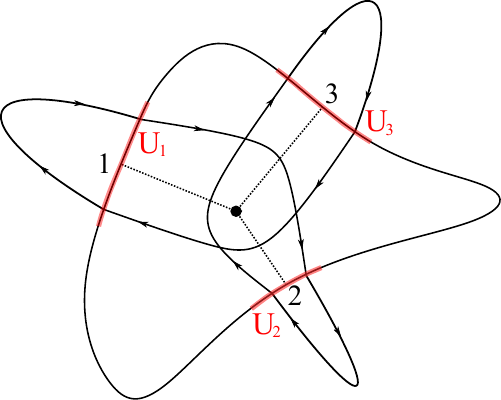}
	\caption{Graphical representation of the projection map $\Pi$ between refraction billiards' trajectories and bi-infinite sequences of symbols. Here, one has $m=3$ nondegenerate central configurations, around which the neighbourhoods $U_i$, $i=1,2,3$ have been constructed. The periodic trajectory displayed realizes the periodic word $\underline s=(\dots,1,2,3,1,2,3,\dots$).  }
	\label{fig:realizes}
\end{figure}
With reference to Figure \ref{fig:realizes}, we proceed as follows: given an admissible domain, suppose that it admits $m$ central configurations $\xi_1,\dots,\xi_m$, where $m\geq2$ by definition. We then construct $U_1,\dots,U_m\subset\partial D$ suitable neighborhoods of $\gamma(\xi_1), \dots,\gamma(\xi_m)$, and say that a trajectory starting with initial conditions $(\xi_0,\alpha_0)$ realizes the word $\underline s\in\{1,\dots,m\}^\Z$ if and only if it crosses $\partial D$ only in $\cup_{i=1}^{m}U_i$ and in the order prescribed by $\underline s$ itself: in such case, we define the projection $\Pi(\xi_0,\alpha_0)=\underline s$. Putting some restrictions on the neighbourhoods $U_i$ as well as on the bi-infinite sequences in $\{1,\dots,m\}^\Z$, one can prove that: 
\begin{itemize}
	\item there exists a set $X$ of initial conditions $(\xi_0,\alpha_0)$ for which $\Pi$ is well defined, that is, the backwards and forwards trajectories with initial conditions $(\xi_0,\alpha_0)\in X$ cross $\partial D$ close to a central configuration (it is said that such trajectories \textit{shadow} the homothetic ones); 
	\item under the hypotheses of nondegeneracy and non-antipodality of central configurations and for $h$ sufficiently large, $\Pi$ is bijective and continuous on $X$, and hence $\mathcal F_{|_X}$ is topologically conjugated to Bernoulli shift, thus chaotic. To prove the surjectivity of the map, we shall use of \textit{Poincar\'e-Miranda theorem} (see \cite{Miranda}), a topological fixed point result, and of the variational characterisation of refracted arcs.  
\end{itemize}
In \cite{barutello2023chaotic}, a thorough analysis on the admissibility condition, including results that still hold when some of the requirements are weakened, is carried on; furthermore, conditions on the bi-infinite sequences which lead to a non-collisional dynamics (namely, whose trajectories do not collide with the central mass) are detected.  
\begin{remark} 
	Although \cite{barutello2023chaotic} is focused on the case of a refractive galactic billiard, it can be proved that a result analogous to Theorem \ref{thm:chaos}, with the same admissibility conditions, holds in the class of \textit{Kepler reflective billiards} . In this case, the chaoticity result represent an interesting complement of the result obtained by Takeuchi and Zhao in \cite{takeuchi2021conformal}. Here, as a consequence of a more general result, they prove that \textit{a Kepler reflective billiard whose boundary is a focused ellipse} (i.e. an ellipse with one of the foci put at the origin, see Figure \ref{fig:ammissibile}, Left) induces an integrable dynamics.
\end{remark}
On one hand, Takeuchi and Zhao's result is completely coherent with Theorem \ref{thm:chaos}, since a focused ellipse is not an admissible domain according to Definition \ref{def:ammissibile} (it indeed has two nondegenerate central configurations, parallel to the $x-$axis, but they are antipodal); on the other hand, the presence of a transition between integrable and chaotic systems by simply translating the boundary is a nontrivial fact that is worthy of a further investigation. In \cite{IreViNota}, we start by proving, analytically, that elliptic (galactic refraction and Kepler reflection) billiards are almost always chaotic, provided the inner energy is sufficiently large. More precisely, fixed a non-circular elliptic shape with any eccentricities, it can be put almost everywhere with respect to the origin (keeping $0$ in its interior) to have an admissible domain.\\ 

We conclude this Section by highlighting how admissibility is only a \textit{sufficient condition} ensuring chaos for large inner energies: it is then natural to ask what happens when such condition is violated but there are no evidences of integrability. In such case, one may ask themselves whether the limit of our proof is only constructive, in the sense that, taking a different projection map, some of the admissibility conditions can be weakened while still obtaining chaos. This will be the subject of a forthcoming paper.  

\section{Conclusions and further perspectives}
 
This paper aimed to show the potential of analytical and semi-analytical techniques when applied to the investigation of dynamical systems coming from Celestial Mechanics, by presenting two examples, the first one of which was on a planetary scale and the second one on a galactic one. \\

The first problem addressed was the search of stability estimates for the motion of a small body moving around the Earth in the so-called geolunisolar model, in the sense that stability times up to which the variation of the orbital elements of a given geocentric orbit have been computed. The results have been obtained by means of two different techniques: the first, based on normal form theory, provided stability times of the order of $10^4$ years for the quantity $\mathcal I=\sqrt{\mu_E a}\sqrt{1-e^2}(1-\cos{i})$, $e$ and $i$ being respectively the eccentricity and the inclination of the orbiting body, which hold for quasi-circular and quasi-equatorial initial trajectories. The second method, based on the celebrated Nekhoroshev theorem, finds its application on a wider set of initial conditions in terms of eccentricities and inclinations; as for the semimajor axis, the results in this case hold for a strip in MEO from 11000 to 19000 km, with stability times that, starting from being very long ($10^5$ years) for low distances, tend to decrease with the altitude. The reason for such worsening in the estimates can be found in the convergence of the constructive algorithm used, and could be potentially improved. On the other hand, another fact which worsen our estimates are given by the presence of lunisolar resonances: in such case, a certain improvement could be achieved by considering Nekhoroshev theorem in its complete form instead of the nonresonant one, and proceeding with a careful analysis of our secular geolunisolar Hamiltonian whenever a resonance occurs. \\
We stress that the approach presented in Section \ref{sec:satelliti} could be potentially used to provide stability estimates in any gravitational system where small masses orbit around a central extensive body, of which the approximated shape (and, as a consequence, the corresponding gravitational potential) is known, even in presence of third bodies which act as a perturbation.\\

As for the second model considered, it has been studied with analytical techniques coming from billiards' and, more generally,  classical dynamical systems' theory. In this case, many results regarding the presence of equilibrium, periodic and quasi-periodic orbits are provided, along with evidences that, under suitable conditions, the central mass acts as a scatterer, deflecting our particle in a chaotic, an somehow unpredictable, way. \\
In general, refraction billiards where two differential potentials are coupled through a refraction interface can be used as a simplified model to study any complex dynamical system whose behaviour presents different regimes, for example depending on the particle's position; in such general case, the same formalism, as well as the same techniques, presented in Section \ref{sec:biliardi} can be used.

\bibliographystyle{apalike}
\bibliography{references}

\begin{thebibliography}{10}

\bibitem{report2022}
{ESA}'s annual space environment report.
\newblock Technical report, European Space Agency, Space Debris Office, 04
  2022.

\bibitem{baldoma2022breakdown}
I.~Baldom{\'a}, M.~Giralt, and M.~Guardia.
\newblock Breakdown of homoclinic orbits to {L}3 in the {RPC3BP} ({I}).
  {C}omplex singularities and the inner equation.
\newblock {\em Advances in Mathematics}, 408:108562, 2022.

\bibitem{baldoma2023breakdown}
I.~Baldom{\'a}, M.~Giralt, and M.~Guardia.
\newblock Breakdown of homoclinic orbits to {L}3 in the {RPC3BP} ({II}). {A}n
  asymptotic formula.
\newblock {\em Advances in Mathematics}, 430:109218, 2023.

\bibitem{BarCanTeranisotropic}
V.~Barutello, G.~M. Canneori, and S.~Terracini.
\newblock Symbolic dynamics for the anisotropic {$N$}-centre problem at
  negative energies.
\newblock {\em Arch. Ration. Mech. Anal.}, 242(3):1749--1834, 2021.

\bibitem{deblasiterracinibarutelloChaotic}
V.~Barutello, I.~De~Blasi, and S.~Terracini.
\newblock Symbolic dynamics and analytical non-integrability for a galactic
  billiard, In preparation, 2022.

\bibitem{IreViNota}
V.~L. Barutello and I.~De~Blasi.
\newblock A note on chaotic billiards with potentials.
\newblock {\em arXiv e-prints}, pages arXiv--2312, 2023.

\bibitem{barutello2023chaotic}
V.~L. Barutello, I.~De~Blasi, and S.~Terracini.
\newblock Chaotic dynamics in refraction galactic billiards.
\newblock {\em Nonlinearity}, 36(8):4209, 2023.

\bibitem{Bol2017}
S.~V. Bolotin.
\newblock Degenerate billiards in celestial mechanics.
\newblock {\em Regul. Chaotic Dyn.}, 22(1):27--53, 2017.

\bibitem{BolNeg}
S.~V. Bolotin and P.~Negrini.
\newblock Regularization and topological entropy for the spatial {$n$}-center
  problem.
\newblock {\em Ergodic Theory Dynam. Systems}, 21(2):383--399, 2001.

\bibitem{boscaggin2021periodic}
A.~Boscaggin, W.~Dambrosio, and G.~Feltrin.
\newblock Periodic solutions to a perturbed relativistic {K}epler problem.
\newblock {\em SIAM Journal on Mathematical Analysis}, 53(5):5813--5834, 2021.

\bibitem{braga2014ring}
F.~Braga-Ribas, B.~Sicardy, J.~Ortiz, C.~Snodgrass, F.~Roques,
  R.~Vieira-Martins, J.~Camargo, M.~Assafin, R.~Duffard, E.~Jehin, et~al.
\newblock A ring system detected around the {C}entaur (10199) {C}hariklo.
\newblock {\em Nature}, 508(7494):72--75, 2014.

\bibitem{breiter1999lunisolar}
S.~Breiter.
\newblock Lunisolar apsidal resonances at low satellite orbits.
\newblock {\em Celestial Mechanics and Dynamical Astronomy}, 74(4):253--274,
  1999.

\bibitem{cellettiIAU}
A.~Celletti.
\newblock {From infinite to finite time stability in Celestial Mechanics and
  Astrodynamics}.
\newblock {\em Astrophys Space Sci}, 368, 2023.

\bibitem{celletti2007kam}
A.~Celletti and L.~Chierchia.
\newblock {\em KAM stability and celestial mechanics}.
\newblock American Mathematical Soc., 2007.

\bibitem{cellettiDeBlasiEft2023nekhoroshev}
A.~Celletti, I.~De~Blasi, and C.~Efthymiopoulos.
\newblock Nekhoroshev estimates for the orbital stability of {E}arth’s
  satellites.
\newblock {\em Celestial Mechanics and Dynamical Astronomy}, 135(2):10, 2023.

\bibitem{celletti2017dynamical}
A.~Celletti, C.~Efthymiopoulos, F.~Gachet, C.~Gale{\c{s}}, and G.~Pucacco.
\newblock Dynamical models and the onset of chaos in space debris.
\newblock {\em International Journal of Non-Linear Mechanics}, 90:147--163,
  2017.

\bibitem{cellettiferrara}
A.~Celletti and L.~Ferrara.
\newblock An application of {N}ekhoroshev theorem to the restricted three-body
  problem.
\newblock {\em Celest. Mech. Dyn. Astr.}, 64:261--272, 1996.

\bibitem{celletti2017analytical}
A.~Celletti, C.~Gale{\c{s}}, G.~Pucacco, and A.~J. Rosengren.
\newblock Analytical development of the lunisolar disturbing function and the
  critical inclination secular resonance.
\newblock {\em Celestial Mechanics and Dynamical Astronomy}, 127(3):259--283,
  2017.

\bibitem{cellettigiorgilli}
A.~Celletti and A.~Giorgilli.
\newblock On the stability of the {L}agrangian points in the spatial restricted
  problem of three bodies.
\newblock {\em Cel. Mech. Dyn. Astr.}, 50:31--58, 1991.

\bibitem{celletti2021reconnecting}
A.~Celletti, G.~Pucacco, and T.~Vartolomei.
\newblock Reconnecting groups of space debris to their parent body through
  proper elements.
\newblock {\em Scientific Reports}, 11(1):22676, 2021.

\bibitem{celletti2022proper}
A.~Celletti, G.~Pucacco, and T.~Vartolomei.
\newblock Proper elements for space debris.
\newblock {\em Celestial Mechanics and Dynamical Astronomy}, 134(2):11, 2022.

\bibitem{chandrasekhar1967ellipsoidal}
S.~Chandrasekhar.
\newblock Ellipsoidal figures of equilibrium—an historical account.
\newblock {\em Communications on Pure and Applied Mathematics}, 20(2):251--265,
  1967.

\bibitem{daquin2016dynamical}
J.~Daquin, A.~J. Rosengren, E.~M. Alessi, F.~Deleflie, G.~B. Valsecchi, and
  A.~Rossi.
\newblock The dynamical structure of the {MEO} region: long-term stability,
  chaos, and transport.
\newblock {\em Celestial Mechanics and Dynamical Astronomy}, 124(4):335--366,
  2016.

\bibitem{deblasi3D}
I.~De~Blasi.
\newblock Chaoticity in three-dimensional galactic refraction billiards, In
  preparation, 2023.

\bibitem{DebCelEft2021satellites}
I.~De~Blasi, A.~Celletti, and C.~Efthymiopoulos.
\newblock Satellites’ orbital stability through normal forms.
\newblock {\em Proceedings of the International Astronomical Union},
  15(S364):146--151, 2021.

\bibitem{de2021semi}
I.~De~Blasi, A.~Celletti, and C.~Efthymiopoulos.
\newblock Semi-analytical estimates for the orbital stability of {E}arth’s
  satellites.
\newblock {\em Journal of Nonlinear Science}, 31(6):1--37, 2021.

\bibitem{deblasiterracinirefraction}
I.~De~Blasi and S.~Terracini.
\newblock Refraction periodic trajectories in central mass galaxies.
\newblock {\em Nonlinear Anal.}, 218:Paper No. 112766, 40, 2022.

\bibitem{deblasiterraciniOnSome}
I.~De~Blasi and S.~Terracini.
\newblock On some refraction billiards.
\newblock {\em Discrete and Continuous Dynamical Systems}, 43(3\&4):1269--1318,
  2023.

\bibitem{Delis20152448}
N.~Delis, C.~Efthymiopoulos, and C.~Kalapotharakos.
\newblock Effective power-law dependence of {L}yapunov exponents on the central
  mass in galaxies.
\newblock {\em Monthly Notices of the Royal Astronomical Society},
  448(3):2448--2468, 2015.

\bibitem{Dev_book}
R.~L. Devaney.
\newblock {\em An introduction to chaotic dynamical systems}.
\newblock Studies in Nonlinearity. Westview Press, Boulder, CO, 2003.
\newblock Reprint of the second (1989) edition.

\bibitem{docarmo2016differential}
M.~P. Do~Carmo.
\newblock {\em Differential geometry of curves and surfaces: revised and
  updated second edition}.
\newblock Courier Dover Publications, 2016.

\bibitem{efthymiopoulos2011canonical}
C.~Efthymiopoulos.
\newblock Canonical perturbation theory; stability and diffusion in
  {H}amiltonian systems: applications in dynamical astronomy.
\newblock In {\em Workshop Series of the Asociacion Argentina de Astronomia},
  volume~3, pages 3--146, 2011.

\bibitem{fasso1989composition}
F.~Fass{\`o} and G.~Benettin.
\newblock Composition of {L}ie transforms with rigorous estimates and
  applications to {H}amiltonian perturbation theory.
\newblock {\em Zeitschrift f{\"u}r angewandte Mathematik und Physik ZAMP},
  40(3):307--329, 1989.

\bibitem{ferrarese2005supermassive}
L.~Ferrarese and H.~Ford.
\newblock Supermassive black holes in galactic nuclei: past, present and future
  research.
\newblock {\em Space Science Reviews}, 116(3-4):523--624, 2005.

\bibitem{froeschle1997fast}
C.~Froeschl{\'e}, E.~Lega, and R.~Gonczi.
\newblock Fast {L}yapunov indicators. application to asteroidal motion.
\newblock {\em Celestial Mechanics and Dynamical Astronomy}, 67(1):41--62,
  1997.

\bibitem{gasiorek2021dynamics}
S.~Gasiorek.
\newblock On the dynamics of inverse magnetic billiards.
\newblock {\em Nonlinearity}, 34(3):1503, 2021.

\bibitem{giorgilli2022notes}
A.~Giorgilli.
\newblock {\em Notes on Hamiltonian dynamical systems}, volume 102.
\newblock Cambridge University Press, 2022.

\bibitem{giorgilliskokos}
A.~Giorgilli and C.~Skokos.
\newblock On the stability of the {T}rojan asteroids.
\newblock {\em Astron. Astrophys.}, 317:254--261, 1997.

\bibitem{gole2001symplectic}
C.~Gol{\'e}.
\newblock {\em Symplectic twist maps: global variational techniques},
  volume~18.
\newblock World Scientific, 2001.

\bibitem{guzzo2023theory}
M.~Guzzo and E.~Lega.
\newblock Theory and applications of fast {L}yapunov indicators to model
  problems of celestial mechanics.
\newblock {\em Celestial Mechanics and Dynamical Astronomy}, 135(4):37, 2023.

\bibitem{hasselblattkatok}
B.~Hasselblatt and A.~Katok.
\newblock {\em A first course in dynamics}.
\newblock Cambridge University Press, New York, 2003.
\newblock With a panorama of recent developments.

\bibitem{devaneyhirschsmale}
M.~W. Hirsch, S.~Smale, and R.~L. Devaney.
\newblock {\em Differential equations, dynamical systems, and an introduction
  to chaos}.
\newblock Elsevier/Academic Press, Amsterdam, third edition, 2013.

\bibitem{kalapotharakos2008rate}
C.~Kalapotharakos.
\newblock The rate of secular evolution in elliptical galaxies with central
  masses.
\newblock {\em Monthly Notices of the Royal Astronomical Society},
  389(4):1709--1721, 2008.

\bibitem{kaloshinsorrentinolocalbirkhoff}
V.~Kaloshin and A.~Sorrentino.
\newblock On the local {B}irkhoff conjecture for convex billiards.
\newblock {\em Ann. of Math. (2)}, 188(1):315--380, 2018.

\bibitem{Kaula1962}
W.~M. Kaula.
\newblock Development of the lunar and solar disturbing functions for a close
  satellite.
\newblock {\em Astron. J.}, 67:300--303, 1962.

\bibitem{kaula1966theory}
W.~M. Kaula.
\newblock Theory of satellite geodesy, {B}laisdell publ.
\newblock {\em Co., Waltham, Mass}, 1966.

\bibitem{KingHele}
D.~G. {King-Hele} and D.~M.~C. {Walker}.
\newblock {Predicting the orbital lifetimes of Earth satellites}.
\newblock {\em Acta Astronautica}, 18:123--131, Jan. 1988.

\bibitem{Kn2002}
A.~Knauf.
\newblock The {$n$}-centre problem of celestial mechanics for large energies.
\newblock {\em J. Eur. Math. Soc. (JEMS)}, 4(1):1--114, 2002.

\bibitem{Kozai}
Y.~{Kozai}.
\newblock {Secular perturbations of asteroids with high inclination and
  eccentricity}.
\newblock {\em Astron. J.}, 67:591--598, Nov. 1962.

\bibitem{lerman2021whispering}
A.~Lerman and V.~Zharnitsky.
\newblock Whispering gallery orbits in sinai oscillator trap.
\newblock {\em Physica D: Nonlinear Phenomena}, 425:132960, 2021.

\bibitem{Levi-Civita}
T.~Levi-Civita.
\newblock Sur la r\'{e}solution qualitative du probl\`eme restreint des trois
  corps.
\newblock {\em Acta Math.}, 30(1):305--327, 1906.

\bibitem{Lidov}
M.~L. {Lidov}.
\newblock {The evolution of orbits of artificial satellites of planets under
  the action of gravitational perturbations of external bodies}.
\newblock {\em planss}, 9:719--759, 1962.

\bibitem{Miranda}
C.~Miranda.
\newblock Un'osservazione su un teorema di {B}rouwer.
\newblock {\em Boll. UMI}, 3:5--7, 1940.

\bibitem{morbidelli2002modern}
A.~Morbidelli.
\newblock {\em Modern celestial mechanics: aspects of solar system dynamics}.
\newblock 2002.

\bibitem{moser1962invariant}
J.~M{\"o}ser.
\newblock On invariant curves of area-preserving mappings of an annulus.
\newblock {\em Nachr. Akad. Wiss. G{\"o}ttingen, II}, pages 1--20, 1962.

\bibitem{murray1999solar}
C.~D. Murray and S.~F. Dermott.
\newblock {\em Solar system dynamics}.
\newblock Cambridge university press, 1999.

\bibitem{nekhoroshev1977exponential}
N.~N. Nekhoroshev.
\newblock An exponential estimate of the time of stability of nearly-integrable
  {H}amiltonian systems.
\newblock {\em Uspekhi Matematicheskikh Nauk}, 32(6):5--66, 1977.

\bibitem{nie2021long}
T.~Nie and P.~Gurfil.
\newblock Long-term evolution of orbital inclination due to third-body
  inclination.
\newblock {\em Celestial Mechanics and Dynamical Astronomy}, 133(1):1--33,
  2021.

\bibitem{panov1994elliptical}
A.~A. Panov.
\newblock Elliptical billiard table with newtonian potential.
\newblock {\em Mathematical Notes}, 55(3):334--334, 1994.

\bibitem{Poschel}
J.~P{\"o}schel.
\newblock Nekhoroshev estimates for quasi-convex {H}amiltonian systems.
\newblock {\em Mathematische Zeitschrift}, 213(1):187--216, 04 1993.

\bibitem{Rosengren2013}
A.~J. Rosengren and D.~J. Scheeres.
\newblock Long-term dynamics of high area-to-mass ratio objects in high-{E}arth
  orbit.
\newblock {\em Adv. Space Res.}, 52:1545--1560, 2013.

\bibitem{rosengren2014classical}
A.~J. Rosengren, D.~J. Scheeres, and J.~W. McMahon.
\newblock The classical laplace plane as a stable disposal orbit for
  geostationary satellites.
\newblock {\em Advances in Space Research}, 53(8):1219--1228, 2014.

\bibitem{Seif}
H.~Seifert.
\newblock Periodische {B}ewegungen mechanischer {S}ysteme.
\newblock {\em Math. Z.}, 51:197--216, 1948.

\bibitem{Shute}
B.~E. {Shute} and J.~{Chiville}.
\newblock {The lunar-solar effect on the orbital lifetimes of artificial
  satellites with highly eccentric orbits}.
\newblock {\em Planetary and Space Science}, 14(4):361--369, Apr. 1966.

\bibitem{steichen1997long}
D.~Steichen and A.~Giorgilli.
\newblock Long time stability for the main problem of artificial satellites.
\newblock {\em Celestial Mechanics and Dynamical Astronomy}, 69(3):317--330,
  1997.

\bibitem{Tabbook}
S.~Tabachnikov.
\newblock {\em Geometry and billiards}, volume~30 of {\em Student Mathematical
  Library}.
\newblock American Mathematical Society, Providence, RI; Mathematics Advanced
  Study Semesters, University Park, PA, 2005.

\bibitem{takeuchi2021conformal}
A.~Takeuchi and L.~Zhao.
\newblock Conformal transformations and integrable mechanical billiards.
\newblock {\em Advances in Mathematics}, 436:109411, 2024.

\end{thebibliography}

\end{document}